\documentclass[12pt, a4paper]{article}

 \usepackage[bottom]{footmisc}

\usepackage{graphicx}
\usepackage{amsthm}   
\usepackage{amsmath} 
\usepackage{amssymb}  
\usepackage{mathrsfs} 
\usepackage{stmaryrd} 
\usepackage{txfonts} 

\usepackage{hyperref}  
\usepackage[capitalize]{cleveref}

\usepackage{enumerate}
\usepackage{appendix} 

\usepackage[all]{xy}
\usepackage[]{bm}

\newcommand{\mathsym}[1]{{}}
\newcommand{\unicode}[1]{{}}

\newcommand{\EEE}{{\mathcal{E}}}
\newcommand{\R}{{\mathbb{R}}}

\newcommand{\N}{{\mathbb{N}}}

\newcommand{\ch}{{\rm{ch}}}
\newcommand{\pa}{{\rm{pa}}}

\usepackage{color}  
\RequirePackage[dvipsnames,usenames]{xcolor}

\usepackage[obeyFinal]{todonotes}
\usepackage{ifthen,xspace}

\newcommand{\eq}[1]{\begin{align}#1\end{align}}

\newcommand{\be}{\begin{equation}}
\newcommand{\bel}[1]{\begin{equation}\label{#1}}
\newcommand{\qe}{\end{equation}}
\newcommand{\ee}{\end{equation}}
\newcommand{\eeq}{\end{equation}}
\newcommand{\ba}{\begin{eqnarray}}
\newcommand{\ea}{\end{eqnarray}}

\def\bal#1\eal{\begin{align}#1\end{align}}
\def\bann#1\eann{\begin{align*}#1\end{align*}}

\date{\today}                      

\begin{document}

\title{The computational power of a human society: a new model of social evolution}

\author{David H. Wolpert \\
 Santa Fe Institute, 1399 Hyde Park Road, Santa Fe, NM, 87501\\
 \texttt{http://davidwolpert.weebly.com}
\\Kyle Harper \\
 University of Oklahoma, 660 Parrington Oval, Norman, OK 73069\\
 Santa Fe Institute, 1399 Hyde Park Road, Santa Fe, NM, 87501\\
 \texttt{http://kyleharper.net}
 }

\maketitle

\begin{abstract}
Social evolutionary theory seeks to explain increases in the scale and complexity of human societies, from origins to present. 
Over the course of the twentieth century, social evolutionary theory largely fell out of favor as a way of investigating human history, 
just when advances in complex systems science and computer science saw the emergence of powerful new conceptions of complex systems, and in particular new methods of measuring complexity.
We propose that these advances in our understanding of complex systems and computer science should be brought to bear on 
our investigations into human history. To that end, we present a new framework for modeling how human societies co-evolve with their biotic environments, recognizing that both a society and its environment are computers.
This leads us to model the dynamics of each of those two systems using the same, new kind of computational machine,
which we define here. For simplicity, we construe a society as a set of interacting
occupations and technologies. Similarly, under such a model, 
a biotic environment is a set of interacting distinct ecological and environmental processes.
This provides novel ways to characterize social complexity, which we hope will cast
new light on the archaeological and historical records. Our framework
also provides a natural way to formalize both the energetic (thermodynamic) costs required by a society  
as it runs, and the ways it can extract thermodynamic resources from the environment in order to pay for those costs --- and perhaps to
grow with any left-over resources. 

\end{abstract}


  \section{Introduction}
\label{sec:introduction}

How and why have human societies grown in scale and complexity over time? When \textit{H. sapiens} emerged
around $300,000$ years ago, 
humans lived in small bands of hunter-gatherers; energy harvest per capita
was of the order of bodily metabolism (i.e. a small multiple of 2000 kC/per capita/per day); 
global population was low ($\sim10^4$); and social roles were relatively unspecialized. The first humans had technology (fire, stone tools) and impressive cognitive abilities.
Today, however, with essentially identical cognitive potential, \textit{H. sapiens}  numbers 8 billion individuals; we consume $\sim10^{4-5}$ kC/per capita per day; and we inhabit globe-spanning social networks that facilitate extreme specialization \cite{morris2010,morris2013measure}.

Social evolutionary theories attempt to explain this growth over time in universal cross-cultural terms \cite{carneiro2018evolutionism,sanderson1999social}. Here, we propose a new framework for social evolutionary history grounded in the theory of computation. Our framework offers distinct advantages. First, by applying the formalisms of computer science theory in the context of human social systems, we can benefit from the rigor of this advanced body of thought about the nature of complexity; a sounder definition of human social complexity might be informed by theories of algorithmic and / or computational complexity. Two, it opens the possibility of a model of social complexity compatible with physics and biology, a longstanding goal of ``big history." Third, given the rich and growing scholarship on the thermodynamics of computation, we can argue for the importance of coupled increases in energy harvesting and information processing as a key factor in human success. Both free energy harvesting and differentiation/specialization have a venerable place in the history of efforts to measure and model social complexity, so our framework is not incompatible with much existing empirical and conceptual work. The result might be a richer model of the dynamics of growth, seen as the interplay of information processing and energy harvesting. 

Our idea is motivated by the success of computation-centered approaches in biology. Already Erwin Schroedinger, several years before the discovery of the DNA molecule, recognized that living systems must maintain themselves away from thermal equilibrium by consuming free energy and processing information (following an ``elaborate code-script," as he put it) \cite{schr44}. He also intuited the
necessity of an ``aperiodic crystal" for the storage of biological information that guides this process. 
In addition, the theory of ``Major Evolutionary Transitions" (METs~\cite{smith1999major,jablonka2006evolution}) in the history of life provides an important paradigm in which radical changes in biological information storage and processing led to major leaps in the complexity of the biosphere. More generally, there have recently been
approaches to theoretical biology grounded in computer science, which are providing
new insights into what life is and how it operates \cite{jost_biological_2020,al-hashimi2023turing}. The modern view is that
organisms and biological systems can be viewed as essentially computational machines, which process inputs to produce outputs. The net result of that processing is forms of biological organization that reliably transmit information to future generations.

In light of this work in evolutionary biology, we view the
emergence of the social systems of modern humans in the Holocene as an ongoing MET. In this current MET, human culture represents a new medium, complementary to DNA, for the accumulation and transmission of information about the external environment. A human society uses information, acquired and stored over the current and previous generations, to process inputs from its environment
and to produce outputs, which are actions that it takes on that environment. In this sense,
a human society is fundamentally a computer. We describe social evolution as the growth of humanity's collective metabolic-computational capacities. We are pursuing Schroedinger's observation that ``living matter, while not eluding the `laws of physics' as established up to date, is likely to involve `other laws of physics' hitherto unknown, which, however, once they have been revealed, will form just as integral a part of this science as the former." Those laws, we believe, require accounting for increasing complexity via the interplay of computation and metabolism 
--- on scales ranging from cells to societies.

What is the ``information'' stored in a human society and how is it used to perform ``computations" on the environment? We think it is intuitive that this information
is represented by the algorithmic instructions that the society runs in order to propagate itself into the future.
More concretely, we propose that computation in social systems is the interaction of social agents with each other and with the environment to harvest energy, to reshape the environment, to reproduce, and to acquire new information. Over time, human societies have grown in scale and complexity because (to borrow terms from the economics literature) the collective ``stock of knowledge" or ``stock of ideas" stored in human culture has grown. 

In essence, we propose to formalize the stock of knowledge as ``information." In this view, 
ultimately it is not our sheer cognitive ability, nor our sociality (both of which are important to human success), that accounts for 
the apparent exceptionalism of  \textit{H. sapiens} in the stream of evolution. Rather it is
our use of symbolic language, computationally powerful enough to support recursion and self-reference.
Sophisticated verbal language, later amplified by information technology like writing and digital computers, allows the accumulation of information over time in culture~\cite{powell_writing_2009,hilbert_worlds_2011}. Culture, here, would represent the totality of the rules, recipes, blueprints, and instructions operated by a society.  

It has been famously difficult to operationalize measurements of complexity, differentiation, and information processing in human societies. While our framework will not suggest any easy answers to this challenge, we will offer a practical stratgy for combining theory, model, and measurement.  We suggest occupational specialization as a proxy for specifying the information stored and processed in human societies. Since Adam Smith, specialization has been acknowledged as a main feature of economic growth. In Smith's words, ``It is the great multiplication of the productions of all the different arts, in consequence of the division of labor, which occasions, in a well governed society, that universal opulence which extends itself to the lowest ranks of the people"~\cite{smith1776inquiry}. An occupation is like an ensemble of algorithms, a set of programs for how to interact with other agents and ultimately the external environment. An occupation is in essence a work tape, encoding instructions for performing computations. Occupational specialization has dramatically increased over the course of human history, as societies have grown and learned to perform a greater variety of operations on the environment. We propose that occupational specialization offers a practical, coarse-grained way to define and measure social complexity in computational terms.


We think that ultimately models of information and computation might also gain by greater dialogue with the social sciences, which present important real-world applications (but resist the sort of clean mathematization favored in textbooks on information theory and computation). If we are right, the road ahead is arduous. But because information processing is so fundamental to the universe, and to the nature of complexity, it is worth trying to reimagine the evolution of human societies as a story in which computation is central.

\subsection{Where we are and where we want to be}

In conventional historical social sciences, or some climate change time-series analysis, or some time-series analysis of epidemics, or even in vast swathes of evolutionary biology and allometric scaling theory, one can write down an extremely flexible set of parameterized equations for the
dynamics of the variables that one is interested in (and can measure). Then one fits the parameters of those equations to a dataset
of observations, to infer the dynamic ``laws" governing the system. At root, such an approach is \textit{phenomenological}, considering observations of how various aspects of the systems (and of human societies in particular) evolve over time, and using statistical analyses to provide insight into this phenomenology.

Needless to say, it will continue to be extremely useful to pursue such approaches further, tailored to investigating
the dynamics of human societies ranging across pre-history (early Holocene) through the modern era.
(Arguably, that is precisely the goal of the ongoing research in the Seshat project~\cite{turchin_quantitative_2018,shin2020scale}.)

However, such a phenomenological approach cannot give us any insight into the other METs not
involving human societies; most if not all deeper understanding arising from applying this approach would
be one-off, not directly transferable to different situations. And perhaps most importantly, there is no sense in which either information theory or computation modeling per se
can play a role in this approach. Under such a phenomenological approach, the only role that the computation performed by a human society
might play is to provide a motivation for a handful of definitions of ``complexity characteristics'' (to use 
the term in the Seshat project, which is distinct from the use of the term in computer science theory). 
Quantifying systems in terms of such complexity characteristics is analogous to characterizing a modern digital 
computer into three levels: strong, moderate or weak. It does not say
anything specific about what computation is actually being performed on that digital computer (much less about what it would mean for a computer to jump from one level to another).

While we think it will be productive to pursue the conventional, phenomenological approach, we also hope to make progress on a fundamentally more challenging modeling exercise. 

Nobody at present has a clear idea of how best to model computational systems outside of the purely
mathematical, non-physical structures considered in CS theory or their direct translations into statistical
physics models. (See~\cite{wolpert2019stochastic} for the former and see~\cite{wolpert2023stochastic}
for further discussion of the latter point.) But at the same time, 
with a few exceptions focusing on biology (work by Gershenson, Zenil and Chaitin in particular comes to mind),
there has never been a concerted effort to try to figure out just how to do that. Certainly it has
never been extensively pursued in the context of deep history.


This essay maps out how such a project might be envisioned. In the next section (Section 2), we briefly discuss existing definitions of social complexity and introduce our own approach, grounded in computational terms. Next (Section 3), motivated by considerations of what data sets are available or potentially available, we propose a novel proxy for the ``computational complexity'' of human societies over time, taking occupational diversity as a metric. In the final five sections (4-8), we introduce a first-principles model of human societies as computers and suggest how we might capture the dynamics of social evolution, considered as the growth of computational-metabolic capacities. 

We emphasize that the goals of this exploratory paper are to stimulate conversation about a research agenda rather than
provide conclusive answers. We would expect pursuing such research might revise or fill out (or replace)
the ideas sketched in preliminary form here.

\section{Framing Social Complexity}
\label{sec:Framing Social Complexity}

What is social complexity? A complete discussion lies beyond the scope of this essay, but we do wish to retrace some steps where social theory could have taken a different path in the 20th century. We will also introduce and discuss a new framework for considering human social complexity. The hope is that a formal
definition of social complexity, along with the associated framework that involves both computation and the 
harvesting of free energy by human societies, could address the deep mysteries of the
growth of human societies in a richer way. 

To begin, we observe that two crucial intellectual developments have unfolded since the middle of the twentieth century, which are rarely observed in conjunction. First, evolutionary models of human history went mostly out of favor. This happened in part because the historical sciences turned away from hopes for unity with the natural sciences~\cite{harper2013culture,morris2022evolutionary}. Second, theoretical computer science (CS) took shape as a scientific field, with major advances in the definition and understanding of complexity~\cite{mitchell2009complexity,sipser1996introduction,arora2009computational,livi08}. While there has been important cross-fertilization between CS, physics, and biology, to date there
has been little overlap between the study of complex systems from a computational perspective
and the evolutionary study of human societies. We will argue that social complexity is a quantity proportionate to a society's computational power.

\subsection{A brief intellectual history of social evolution}

In large part, this lack of contact between these fields is due to the configuration of academic disciplines. When the modern research university took shape in the late 19th century, an enduring division of labor took hold in the way the past is allotted to different fields~\cite{smail2005grip}. The pre-human past was allotted to scientists (e.g. geologists, paleontologists, and biologists). The prehistoric human past was assigned to anthropologists, who would specialize in the study of “primitive” societies either through ethnography or archaeology. Meanwhile, the last few thousand years – the period of governments and writing – became the domain of historians. This arbitrary division still affects the study of social evolution.

Stadial models of social evolution were developed already in the Enlightenment - by John Millar, Adam Smith, and the Marquis de Condorcet, among others. Herbert Spencer linked stadial history to biological evolution (and highlighted the importance of differentiation as an index of complexity). In its early decades, anthropology was dominated by social-evolutionary models~\cite{tylor1881anthropology,morgan1877ancient}, which proposed a progression from simple or primitive society to more advanced civilization. While value-laden models of ``progress" were adopted by early anthropologists such as E.B. Tylor and Lewis Henry Morgan, subsequent generations of anthropologists reacted against existing stadial models of social evolution. There were, to be sure, western imperialist biases baked into this normative, progressive model of social evolution, which was often influenced by crude social Darwinism~\cite{kuper1988invention}. In reaction, Franz Boas and his influential students proposed relativist definitions of culture. Stadial models, ideas of progress, and evolutionary models came to seem hopelessly tainted with cultural prejudice.

The anti-evolutionary stance was then reinforced by the hermeneutic methods that became predominant in anthropology and the humanities more generally. It would be hard to overstate the influence of the so-called ``linguistic turn'' in professional academic fields such as anthropology and history. The idea that the study of the humanities was a pursuit apart from the study of nature was not new (at the beginning of the century, Wilhelm Dilthey had argued for methods specific to Geisteswissenschaften apart from the Naturwissenschaften). But the linguistic turn cemented the idea that the humanities were concerned with the interpretation of each culture in all its particularity. In this view, developed in the work of Clifford Geertz, human culture is a web of meaning, and the objective of anthropology or history is to interpret culture -- to understand rather than to explain sequences of cultural development over time~\cite{geertz_interpretation_1973}. Qualitative methods were ascendant over quantitative ones, while paradigms drawing inspiration from the natural sciences were branded as ``reductionist" ~\cite{harper2013culture}. 

As has been pointed out (\cite{morris2010}, among others), this work criticizing social evolution does nothing to invalidate the possibility of an objective and neutral measurement of social complexity. A more complex society is not intrinsically ``higher" or ``better." But it is more complex (as objectively, we would argue, as one algorithm or computational problem can be more complex than another, without any necessary moral hierarchy). While research on the dynamics of social complexity has survived in pockets of anthropology, in the discipline of academic history, formal, mathematical frameworks were banished with almost religious zeal. Over the second half of the twentieth century, academic history, as a field, came to prefer qualitative over quantitative methods, to reject scientific models of explanation (on the ground that they are reductionist), and to focus on particularities over generalities. As Ian Morris has noted, whereas most social-science fields maintain a healthy tension between interpretive and explanatory approaches, the commitment of historians “to humanistic and particularistic questions and methods verged on monomaniacal"~\cite{morris2022evolutionary}.

As a lingering consequence of the 19th-century division of labor, anthropologists still tend to study relatively small-scale, simple societies, and this is especially true of anthropologists who are most open to evolutionary thinking. ``Cultural evolutionists" in contemporary anthropology use rigorous evolutionary models to help illuminate the ways in which culture is adaptive and might be transmitted; in particular, ``cumulative cultural evolution" helps to explain how foraging societies can solve complicated problems via the accumulation and transmission of knowledge, stored in culture~\cite{henrich_demography_2004,smaldino2013,boyd_different_2018}. Historians, by contrast, mostly study the last few thousand years of the past, and since evolutionary models are effectively taboo, there are few evolutionary models that try to account for the spectacular growth of human societies in the late Holocene. The net result is that evolutionary models of human history have been nurtured mostly on the margins of anthropology, or within other fields altogether (such as economic history, or the ``big history" which is mostly written by physicists and biologists), or among dissidents like Morris.

Thus, someone naively interested in how human societies have grown in scale and complexity would be surprised to learn that there does not exist a terribly robust literature on what social complexity is or how we have achieved it. There are of course exceptions. Even as Boasian relativism was ascendant, the anthropologist Leslie White sought to reformulate an evolutionary model on more neutral grounds, centered on energy (his model is focused on the following ``equation": culture = energy times technology)~\cite{white_evolution_1959}. Others have adopted the approach of Elman Service, who developed a definition of social complexity based on ``levels of hierarchy," (e.g. band, tribe, chiefdom, state)~\cite{service_cultural_1971}. And over the years, there have always been some (Raoull Naroll, Robert Carneiro, George Murdock, etc.) who have kept alive efforts to define and quantify social complexity, sometimes blending differentiation/specialization, hierarchy, and scale (e.g. the size of the largest settlement)~\cite{naroll1956preliminary,carneiro1967relationship,murdock_measurement_1973,chick_cultural_1997,denton_cultural_2004,gedeon_social_2018}.

The paradigm of Big History addresses the question of social complexity and very often focuses on energy capture and exchange~\cite{christian_maps_2004,spier_big_2015}. However, it has so far not deeply engaged with informational or computational measures of system behavior, or models of dynamics grounded in information theory and / or CS theory (though see~\cite{christian_complexity_2017} for thoughts in this direction). A parallel effort is represented by cliodynamics. Cliodynamics has inspired an ambitious research agenda around social complexity, notably the Seshat project~\cite{turchin_quantitative_2018}. While most of its practitioners are amenable to evolutionary models, cliodynamics has so far not engaged extensively with CS theory or information theory. Most work in this vein, moreover, has so far focused on the development of agrarian societies; the Seshat data go as late as 1900 in some cases, but it is fair to say that the dataset is not designed to offer insights into the dramatic increases in the complexity characteristics of modern industrial and post-industrial societies. In many world regions, maximal complexity was reached hundreds or thousands of years ago. 

Economics has arguably been the most successful branch of social science in addressing big-picture questions about growth and development quantitatively, and here there are many potential points of engagement. First, there is the literature on economic complexity, which, while not historical in its orientation, provides rich models and datasets for contemporary economies~\cite{hausmann_atlas_2013}. More historically, the ``new institutional economics" highlights the importance of (formal and informal) rules in shaping exchange and production~\cite{north_institutions_1990}. We would simply translate these insights into the terms of CS theory, to say that institutions or rules are part of the way that agents interact to process information. Much work in growth theory has underscored the importance of ideas and innovations in increasing productivity. Joel Mokyr, for instance, traces the breakthrough to modern growth back to the spread of practical, empirical science~\cite{mokyr_lever_1990}. ``Unified growth theory" focuses on the dynamical feedbacks that encourage the formation of human capital and leads to the economic-demographic transition from a Malthusian regime to a regime of growth~\cite{galor_unified_2011}. Paul Romer won a Nobel Prize for his seminal contribution to endogenous growth theory, the ``nonrivalry of ideas"~\cite{romer_endogenous_1990}. Unlike other goods that contribute to productivity, ideas are not diminished in being replicated. The Haber-Bosch process, for example, is based on a chemical reaction for synthesizing reactive nitrogen; it has contributed as much or more to human well-being as any innovation, and yet it is a simple recipe or idea that once discovered could be disseminated and replicated easily (relative to its original discovery). Most real growth can be attributed to increases in such a stock of knowledge or ideas. We see our proposal as compatible with these versions of growth theory, but we seek to formalize the nebulous ``stock of knowledge" or ``ideas" as information.

\subsection{Computational approaches to complexity in biology}
\label{sec:computational_approaches}

One consequence of the separation of the historical sciences from the natural sciences is that the study of human social evolution has been largely out of touch with developments in the field of complex systems science. Social evolutionary models have always tended to operate with an informal version of what ``complexity" is - taking a ``you know it when you see it" approach and sticking to proxies like specialization, hierarchy, and scale. This really bears underscoring. Even among those who think explicitly most about social complexity, there is very little effort to define complexity formally, much less in terms that are independent of the domain of human societies. But from the 1970s on, important developments have helped clarify what complexity means. A complex system is characterized by the interaction of parts, such that the dynamical properties of the system are emergent from the interaction itself rather than reducible to the properties of the system's subcomponents. It should be emphasized that there is no consensus definition or model of complexity. However, much progress has focused on a cluster of related ideas centered around algorithmic complexity (sometimes called ``Kolmogorov complexity," after the Soviet scholar Andrey Kolmogorov) and computational complexity. 

First,  CS theory
developed the concept of the ``algorithmic (Kolmogorov) complexity of an object (represented as a bit string)'',
defining it as the 
shortest program one could write in some fixed programming language like Python that would reproduce that
object and then halt. Algorithmic complexity provides a way to formally capture how much effort needs to be expended in providing
instructions to a computer for how to construct an object. 

Complementing this kind of complexity, computational complexity
considers how difficult the actual construction process is~\cite{moore2011nature,mitchell2009complexity,sipser1996introduction,arora2009computational,livi08}. 
Whereas algorithmic complexity concerns an individual
object (bit string), computational complexity involves \textit{problems}, which are defined as (usually) an infinite
set of questions of the form, ``What is the optimal value of \{...\} in situation \{...\}?'', using some pre-defined notion of ``optimal''.
A famous example is the ``traveling salesman'' problem (TSP), defined as all instances of the question, ``What is the shortest
possible route a salesman could follow that connects a set of cities at the following coordinates?''~\cite{moore2011nature,sipser1996introduction,mitchell2009complexity,arora2009computational}. 
Computational complexity asks, ``As the number of bits it takes to specify the situation grows,
how fast must the resources required by a program to answer the associated question increase?'' As an example, 
if we take ``resources'' to mean the time the program takes, computational complexity
would consider questions like, ``As I increase the number of cities in a TSP question, how much extra time must
be taken by any program to answer the question correctly?''

While these new ideas of complexity have left little (if any) impact on the study of human social evolution, they have had a deep impact in evolutionary biology. The discovery of DNA in the 1950s confirmed Schroedinger's intuition, and the flowering of genetics has underscored that information is central to life. Biological inheritance is the transmission of information about the environment and how to do computations in the environment in a molecular substrate. In the words of Richard Dawkins, ``what lies at the heart of every living thing is not a fire, not warm breath, not a spark of life… it is information, words, instructions"~\cite{dawkins_blind_1996}. Selection ultimately acts on information~\cite{adami_use_2012}. The potential of information-theoretic notions of complexity inspired Eörs Szathmáry and John Maynard Smith's influential model of METs~\cite{szathmary_major_1995,szathmary_toward_2015}. These transitions are characterized by changes in the way that biological information is stored, exchanged, and transmitted. The canonical METs includes the emergence of life, protocells, genes, the eukaryotic cell, plastids, multicellularity, eusociality, and human language. The precise list is debated and debatable, but the concept is invaluable, foregrounding the interplay of metabolic and computational capacities in the history of life. The MET framework, with information processing at its center, undergirds most serious attempts to think about complexity in biological evolution. 

The eukaryotic cell - arguably the most important development in evolution after the emergence of life itself, and the basis for all complex organisms - is a perfect example. In its mature form, the eukaryotic cell combines revolutionary increases in information storage and processing with much more powerful metabolic systems (epitomized by the mitochondrion). This is illustrated
by the leap in power harnessed per unit of information in the transition from prokaryotes to eukaryotes (Fig. 1).

\begin{figure}[h]
\centering
\includegraphics[width=6in]{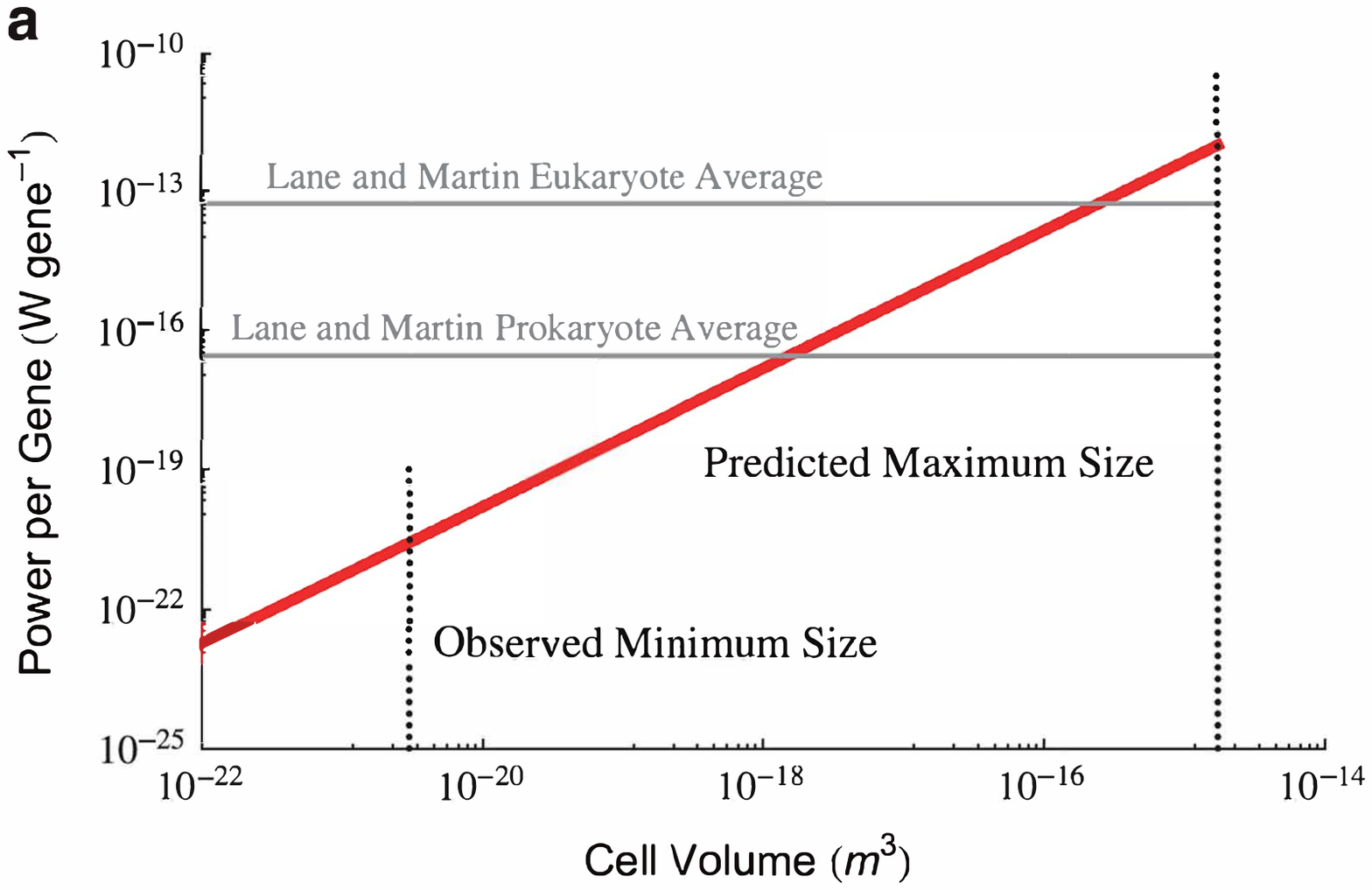}
\caption{Scaling of metabolic rate to cell volume. The y-axis represents the power harnessed per unit of information, in watts per gene, with prokaryote and eukaryote averages in grey lines. Figure from ~\cite{kempes_evolutionary_2016}.} 
\label{fig:1} 
\end{figure}

The computation-centered view of biology helps to explain what is, from a physics perspective, the central conundrum of life: how, in a universe dominated by the Second Law of Thermodynamics, living systems reduce local entropy, creating order and organization~\cite{goldenfeld_life_2011}. The growth of the biosphere is driven by coupled increases in the metabolic-computational capacities of living systems; evolution is a kind  of ``Maxwell's demon" sieving for information that can exploit free energy to reproduce itself~\cite{krakauer_darwinian_2011}. The information stored in living systems is information about the external environment. It is ``how-to" information, ensembles of algorithms for how to self-assemble, how to observe the external environment, how to harvest free energy, how to act in and modify the environment, how to interact with other agents (conspecifics, predators, prey, etc.), and how to reproduce. For example, oxygenic photosynthesis, from this perspective, is an algorithm, a supremely successful set of step-by-step instructions for operating a chemical transformation that turns carbon dioxide, water, and sunlight into sugar. The ``idea" of photosynthesis can be translated into chemical or even binary symbols ($6CO_2 + 6H_2O \rightarrow  C_6H_{12}O_6 + 6O_{24}$). Of course in plants this program is operated by a network of organelles, enzymes, and genetic material, but it may indeed offer more ``bang per bit" than any other genetic algorithm on Earth. 

Needless to say, informational and computational approaches to biology are not always appropriate or useful at a given level of investigation. But they are truly fundamental, related to the deep connection between the rules of the living world and the physical universe. They help to address the biggest questions about the expansion of life. The ``other laws of physics" that Schroedinger foresaw are the laws of complexity - of emergence and interaction, shaped and constrained by the interplay of metabolism and computation. We propose that these other laws of physics be extended to human social evolution. 

Just as the long-run growth of the biosphere (measured in biomass, energy harvesting and exchange, and complexity) is a central question of macroevolution, so the long-run growth of human societies is the central question of human social evolution. We agree with Maynard Smith and Szathmary that the emergence of humans marks a MET and that language is central to this transition. Yet it is not just the emergence of our species that requires exploration in these terms. The last 300K years of human history can be considered an ongoing MET. We take growth - in scale, metabolism, and computational complexity - as the proper explanandum social evolutionary theory. 

In the sequel we present a mathematical definition of a new kind of computational machine,
and our framework for modeling the co-evolution of a society and its environment as the joint dynamics
of two of these machines.
We had several goals when constructing this framework, which in some senses were at cross-purposes. Foremost, we want 
the framework to
be well-suited to investigating the issues described. In addition though we want it to be possible (or at least conceivable)
to gain insight into its behavior using purely mathematical analysis, e.g., by appropriately extending CS theory. 
We also tried to ensure that
the framework can be gainfully applied to investigate multiple METs, not just the current one.
These goals drove us to try to formulate as simple a framework as possible, i.e., as ``coarse-grained'' a framework
as we could. On the other hand, another goal we had was that it be possible to readily tailor the framework 
to the special case of the development of human societies. In particular, we want it to be possible to
tailor the framework to incorporate the kinds of data sets we either now have or will in the near future.
This drove us to try to formulate as rich a framework as possible, i.e., as ``fine-grained'' a framework as we could.
Trying to meet these opposing goals led us to a framework that is ``moderate-grained'', in the same sense as the frameworks
in~\cite{west1997general,west2001general,enquist1998allometric,bettencourt2013origins,bettencourt2020urban,arroyo2022general}


 \section{Measuring Computational Complexity in Human Societies: A New Proxy}
\label{sec:Measuring Complexity in Human Societies: A New Proxy}

There is a straightforward reason why energy harvesting has been an appealing index of social evolution. Energy harvesting can be measured in objective, quantifiable units (calories or joules) which are physically constant over time and space, permitting direct comparison: 500 kC of antelope meat on the African savanna 100Kya can be directly compared with 500kC of quinoa salad in a Santa Fe bistro in 2023. But what is measurable is not necessarily what is important - or in this case, is not the whole story. Even though information is also a fundamentally physical entity, quantifiable in bits, it is much more challenging in practice to measure information storage and processing outside of simple systems or mechanical computing devices. It is far from clear how much computation is occurring even in simple cellular operations that can be directly observed in laboratory conditions. Measuring information and computation in human societies is a daunting challenge. For past societies, yet more challenging still.

Various proxies for information storage and processing in human societies could be envisioned, and it is ultimately a question of both practical and theoretical considerations how best to approach the flow of information over time in human societies. 
For instance, we might just measure the scale of the ``datasphere,'' the number of bits stored in extra-bodily media such as books and memory chips (in
2025, it is estimated globally to be 175 zetabytes). While this is not uninteresting, obviously many of these bits are inconsequential - think cat videos rather than the Haber-Bosch process. To put it formally, they are ``syntactic" (independent of meaning) rather than ``semantic'' (meaningful, in the sense of causally effective: see ~\cite{kolchinsky_semantic_2018}). In the pithy formulation of Artemy Kolchinsky and David Wolpert, they have little ``bang per bit." A better measure has been proposed by Ian Morris, whose Social Development Index includes an Information Technology score (which takes into account literacy rates, writing, information storage, and communication technology) on a scale from 1-100~\cite{morris2010}. While this is still admittedly ``crude'' and not paired to any formal theory of information, it has the distinct virtue of making a start of things. 


Our goal is to develop a proxy that best aligns with the theoretical framework we are proposing. In Section 4, we introduce a formal model 
of human societies as hierarchies of interacting computational machines, where each separate machine 
can be identified in many different ways as components of real human societies. As an example, 
we could identify each machine with a separate human. This would be particularly appropriate for analyzing small
bands of hunter-gatherers. Another possibility, appropriate for analyzing larger Neolithic societies, would be 
to identify each machine as a separate city in a region, e.g., ancient Mesopotamia. Alternatively, focusing within
a single such city, we could identify each machine as a separate occupation, and / or technology. Moving up to the
present day, we could identify each machine as a separate firm in a modern state's economy. 

Note that in all of these examples, the precise set of machines will vary over time, and one of the major ``tasks'' of the overall system
of interacting machines is to extract energy from the physical environment, both to feed the members of the
associated society, and for the society to use as goes about its activities --- a key one of which is extracting
yet more energy from the environment.


\subsection{Occupations as (computing) machines}

To choose among these (and other) choices for how to identify the separate machines in a (computational model of a)
human society we also need to acquiesce to the reality of what data are available or potentially available.
Accordingly, as a novel proxy for the computational power of a human society, we propose that occupational specialization is promising. (See~\cite{hausmann_atlas_2013}, who in a similar spirit suggest thinking of the knowledge or expertise employed by an individual in terms of ``personbytes"). An occupation is a job or a profession. It is a social role centered on labor, irrespective of the extent to which that labor is implicated in market exchange. Occupations can be more or less specialized and require higher or lower degrees of skill and/or training (i.e. from an information-theoretic perspective, not all occupations are equally complex, which means that it would be desirable to do more than simply count occupations to account for the computation going on inside an occupation). Since Adam Smith, the division of labor has been recognized as a hallmark of economic growth. Smith emphasized that the division of labor is limited by the extent of the market, but as Gary Becker and Kevin Murphy long ago observed, specialization is also constrained by coordination costs and --- crucially --- the total stock of knowledge~\cite{becker_division_1992}. The last is a particularly profound insight, although remarkably it has had little influence. We wish to build on it by leveraging the fact that occupational specialization therefore reflects the stock of knowledge. \textit{Ceteris paribus}, the occupations that exist in a given society embody the totality of what a society knows how to do. We also immediately confess that one could make this proxy arbitrarily more sophisticated - and complicated - by considering technologies as computing machines too, since the integration of technology into human societies both replaces occupations and creates new ones. For the first step, we set this issue to one side, a concession to practicality rather than a claim that it is ultimately appropriate.

Occupational specialization reflects both diversification (the expansion of the number of things that a society knows how to do) and division of labor (the finer-grained separation of functions and tasks in an economy, as in Smith's pin factory). Occupational specialization offers the potential to get at the total information stored in a culture and the processing of that information, i.e. computation. The importance that we assign to occupational specialization as an index is compatible with theories of economic growth that emphasize the total stock of knowledge or ideas (ultimately turned into innovations) as the most fundamental cause of long-term gains in productivity (e.g.~\cite{easterlin_growth_1998,jones2002sources,koyama_how_2022}). 

In this view, advances in the Scientific Revolution (e.g. Newtonian mechanics) were ultimately translated by engineers and entrepreneurs into useful productivity-enhancing technological innovations like steam engines (and there are extensive literatures on how discoveries become innovations, who mediates this process, the extent to which innovation is evolutionary, etc.). The Industrial Revolution was transformational not because coal was suddenly available (it was always there), but rather because societies learned how to exploit fossil energy by creating blueprints of steam engines that transformed the mining, textile, manufacturing, and transportation sectors. The Industrial Revolution was, in our terms, a revolution that coupled increases in information processing and energy harvesting; for the first time in our history, these increases were of a speed and magnitude to permit escape from the Malthusian trap. The Second Industrial Revolution (ca. 1880-1930) was spurred by basic discoveries in chemistry and electromagnetism, which eventually produced massive innovation (electricity, light bulbs, telecommunications, internal combustion engines, fertilizer synthesis, polymers, etc.) that spread to scale worldwide over the 20th century. In recent decades, the mass production of microprocessors has been arguably the most important innovation. 

We imagine that the aggregate stock of information encompasses the basic science (e.g. Maxwell's equations) and practical implementation (silicon is a good material for circuits); this aggregate stock of information is then reflected in the composition of an economy's occupations (e.g. research chemist, software programmer). As Hayek recognized, perhaps the most amazing thing that economies do is to coordinate vast amounts of knowledge, even though individual agents themselves only have access to a tiny proportion of that knowledge~\cite{hayek2013use}.
For these reasons, as a first approximation, we propose that the number of distinct occupations necessary to maintain and reproduce a society is a reasonable measure of a society's complexity. 

In our formal model (presented in the following sections) each occupation is considered a distinct machine. A society is considered a computational agent comprised of multiple interacting machines. Any human society is running numerous distinct algorithms simultaneously at each timestep. The reason why occupational specialization is appealing for our purposes is straightforward: an occupation can be considered an ensemble of algorithms for interacting with the environment and with other social agents. 














Categorizing occupations, however, is no trivial undertaking. Several categorizations of occupational specializations already exist~\cite{smith_foundational_nodate}. Each of these schema is organized hierarchically, with varying resolution (e.g. the U.N. scheme recognizes 436 occupations at the most detailed level, the E.U. scheme 3008); each also includes a description of the tasks or skills involved in each occupation. There is also a classification known as HISCO (Historical International Standard of Classification of Occupations), developed by researchers interested in the history of occupations~\cite{leeuwen_hisco_2002}. Obviously, a simple count of occupations will be sensitive to the grain of resolution used to divide up occupations (e.g. engineer, chemical engineer, fertilizer synthesis engineer, and so on). But this is not cause for despair. First, it has already been shown by researchers interested in the scaling patterns of occupational diversity in modern cities that scaling patterns are independent of the degree of resolution~\cite{bettencourt_professional_2014}. Second, it would be possible to use word-embedding tools to identify the actual skills, functions, and tasks underlying occupational titles and to construct measures of the real distances in the networks of skills, functions, and tasks embodied in a given society's occupations. 
%
%

We immediately see non-trivial ways in which this measurement might be too limited even on its own terms. First, as mentioned, not all occupations are created equal, and some occupations are presumably more computationally complex than others. A nuclear physicist may execute more algorithms, and more complex algorithms, than a beet farmer. It would be fruitful to explore this possibility by considering proxies for the computational complexity of a given occupation, such as years of training required. Second, two identical sets of occupations might interact in different ways, permitting meaningfully more or less complex computations. Therefore, the density and nature of interactions between occupations are important, as are the network structures that reflect how occupations communicate. Third, different algorithms, singly or in combination, may offer greater ``bang per bit." In other words, some algorithms (or some occupations) may have a disproportionate thermodynamic effect on a society.

\subsection{Specialization, scaling, and change}

We observe two stylized facts that support the use of occupational specialization as a proxy. First, in the present, occupational specialization is strongly correlated with per capita wealth, underscoring the link between productivity and specialization (Figure 2).

\begin{figure}[h]
   \vglue-2cm
\advance\leftskip-1cm
  	\includegraphics[width=1.25\linewidth]{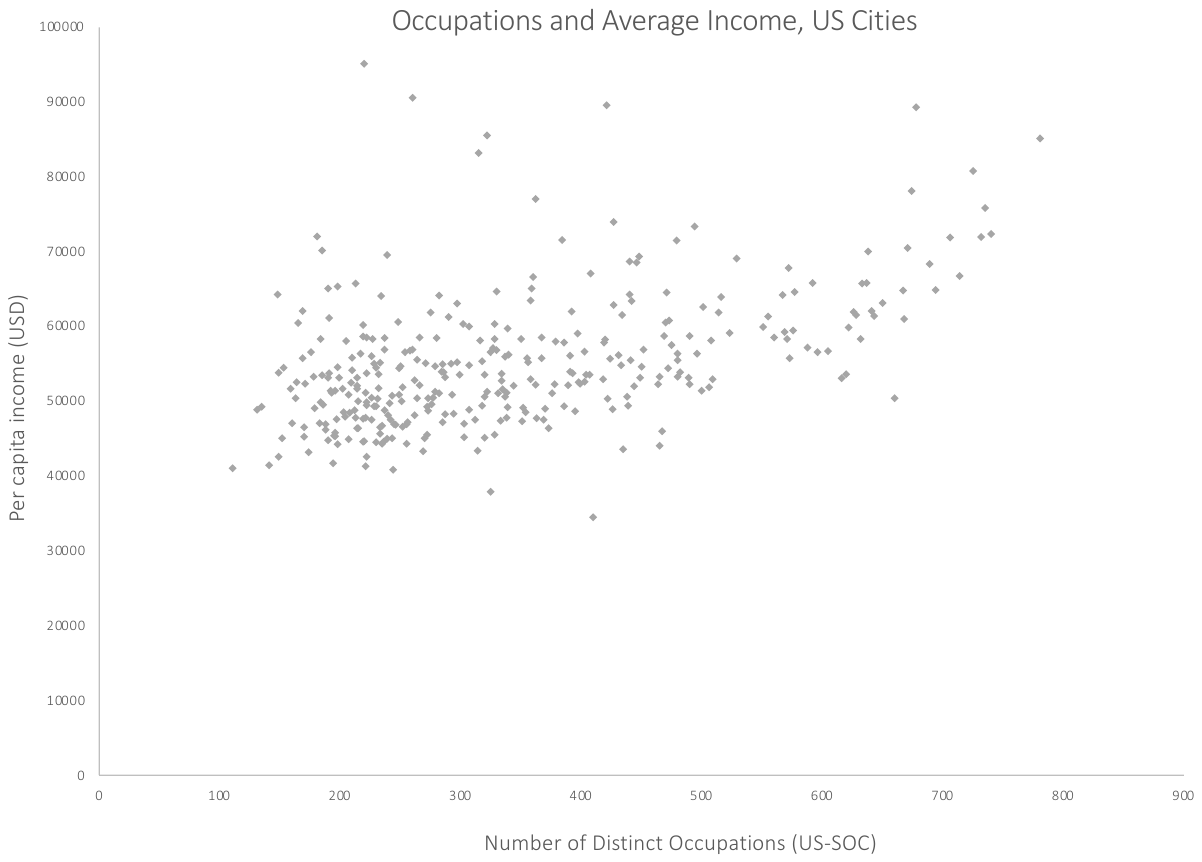}
 	\vglue -10mm
	\hglue 25mm	
\caption{Average income rises with occupational specialization. US Metropolitan Statistical Areas, 2021; three outliers (Odessa, San Jose, Midland) omitted; BLS data, which uses the US SOC (Standard Occupational Classification) 2018 version, with 867 occupations at the most granular level.} 
\label{fig:2} 
\end{figure}

Second, occupational specialization increases over time, mirroring other measures of social complexity such as largest city-size, levels of hierarchy, etc. We also believe that our proxy has the potential to capture recent (last 200 years) increases in social complexity in a more fine-grained way than existing measurements. Consider, as a first approximation, the number of unique words used by census respondents to describe their occupation between 1850 and 1940 in U.S. cities (Figure 3). 

\begin{figure}
\centering
\includegraphics[width=6in]{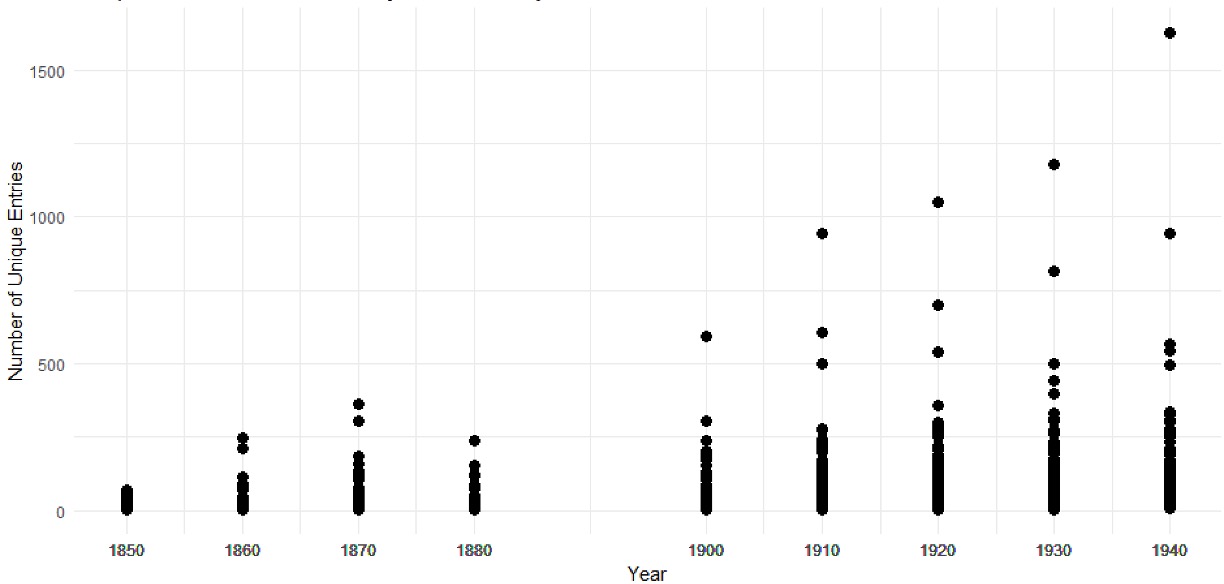}
\caption{Number of unique words (minus articles, prepositions) used by census respondents in US cities to describe occupation; each dot is a US city. Data source: IPUMS (Integrated Public Use Microdata Series). This measure is independent of any classification system, and relies only on the respondent's direct answers, here in a one percent sample.} 
\label{fig:3} 
\end{figure}

Cities are ``social reactors'', ~\cite{bettencourt2013origins} creating dense networks for the flow of information. It has been observed that most socio-economic metrics of modern cities (from income to creativity to crime) scale according to a power law, whereas occupational diversity scales logarithmically, ``illustrating the crucial role that diversity has in fostering a strong economy"~\cite{bettencourt_professional_2014,yang_scaling_2022}. This is an important insight, and indeed it suggests that occupational diversity is a critical ingredient in the growth of cities and the complexity of human societies more generally. In turn, occupational specialization is constrained by various factors, including the extent of the market and coordination costs (both of which are shaped by variables such as institutions, technology, and transportation infrastructure that are highly pertinent to understanding how a society processes information), in addition to what interests us most immediately, the stock of information itself (i.e. the stock of ideas or knowledge in the economist's terms).  

Consider the observation that occupational diversity scales logarithmically with urban population size (Figure 4, from IPUMS, data via the American Community Survey's Public Use Microdata Sample, a classification scheme that contains 530 unique occupations; see also ~\cite{bettencourt_professional_2014}).

\begin{figure}[h]

\centering
\includegraphics[width=6in]{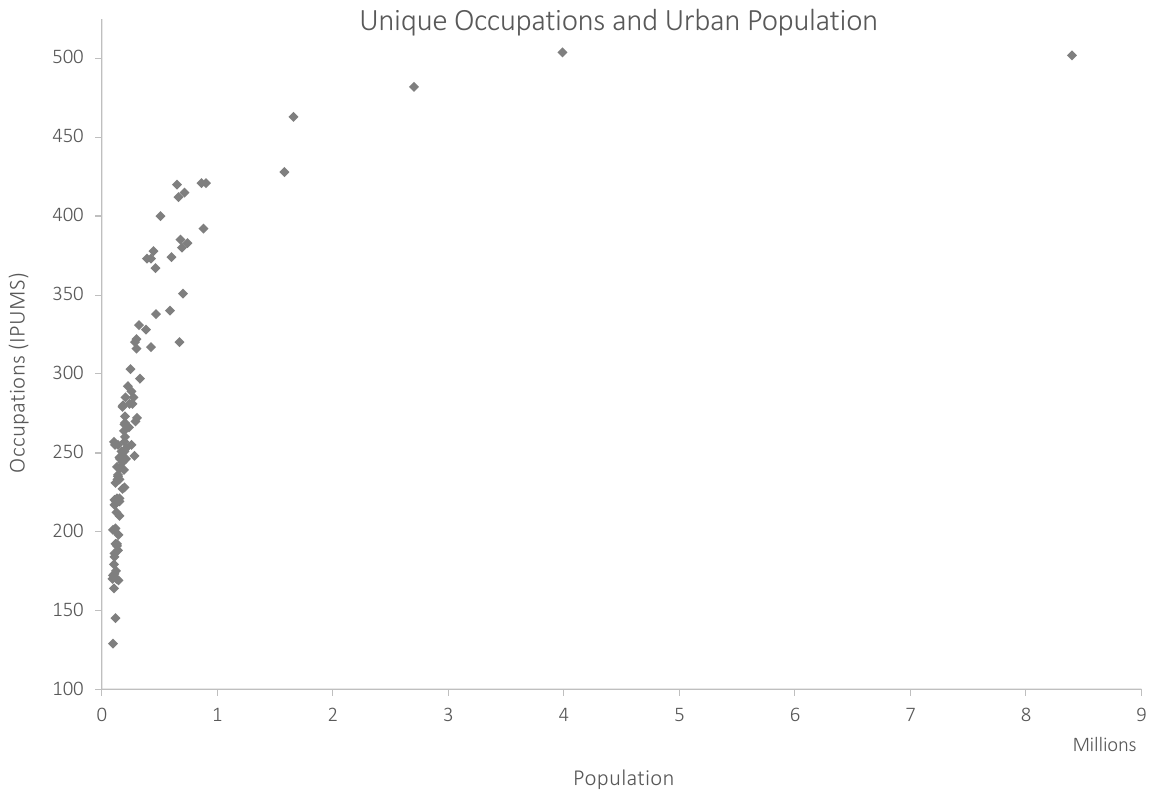}
\caption{Urban populations (millions) and unique occupations, US cities, present (data source: IPUMS, based on the American Community Survey occupation classes).} 
\label{fig:4} 
\end{figure}

Even a rough evaluation of the relationship between occupational diversity and urban population over time suggests interesting variation, however. Occupational diversity in US cities was distinctly lower in the 1940s (in the aftermath of the Second Industrial Revolution) than at present, and lower still in the 1850s (in the midst of the First Industrial Revolution). In preindustrial times, specialization was even more constrained. It is not simply that urban populations were smaller in the past. For instance, if ancient Rome existed in today's urban network, with its population of 1M, we would predict it to have ca. 400 unique occupations, whereas in fact it has fewer than half of that (based on a count of occupational designations in inscriptions and texts from the Roman Empire; see also~\cite{hanson_urbanism_2017,kase_division_2022}). To some extent, these differences are sensitive to the difficulty of maintaining occupational classification at consistent resolution across time and space. But not entirely. Ancient Rome did not have chemical engineers, software developers, flight attendants, web designers, etc., and even at infinite resolution, real shifts have occurred because the stock of information has grown. We hypothesize (as in Figure 5) that the relationship between occupational specialization and population size has changed as a function of social evolution - that the computational-metabolic capacities of human societies have grown due to increases in the stock of information.

\begin{figure}[h]
   \vglue-2cm
\advance\leftskip-2cm
  	\includegraphics[width=1.4\linewidth]{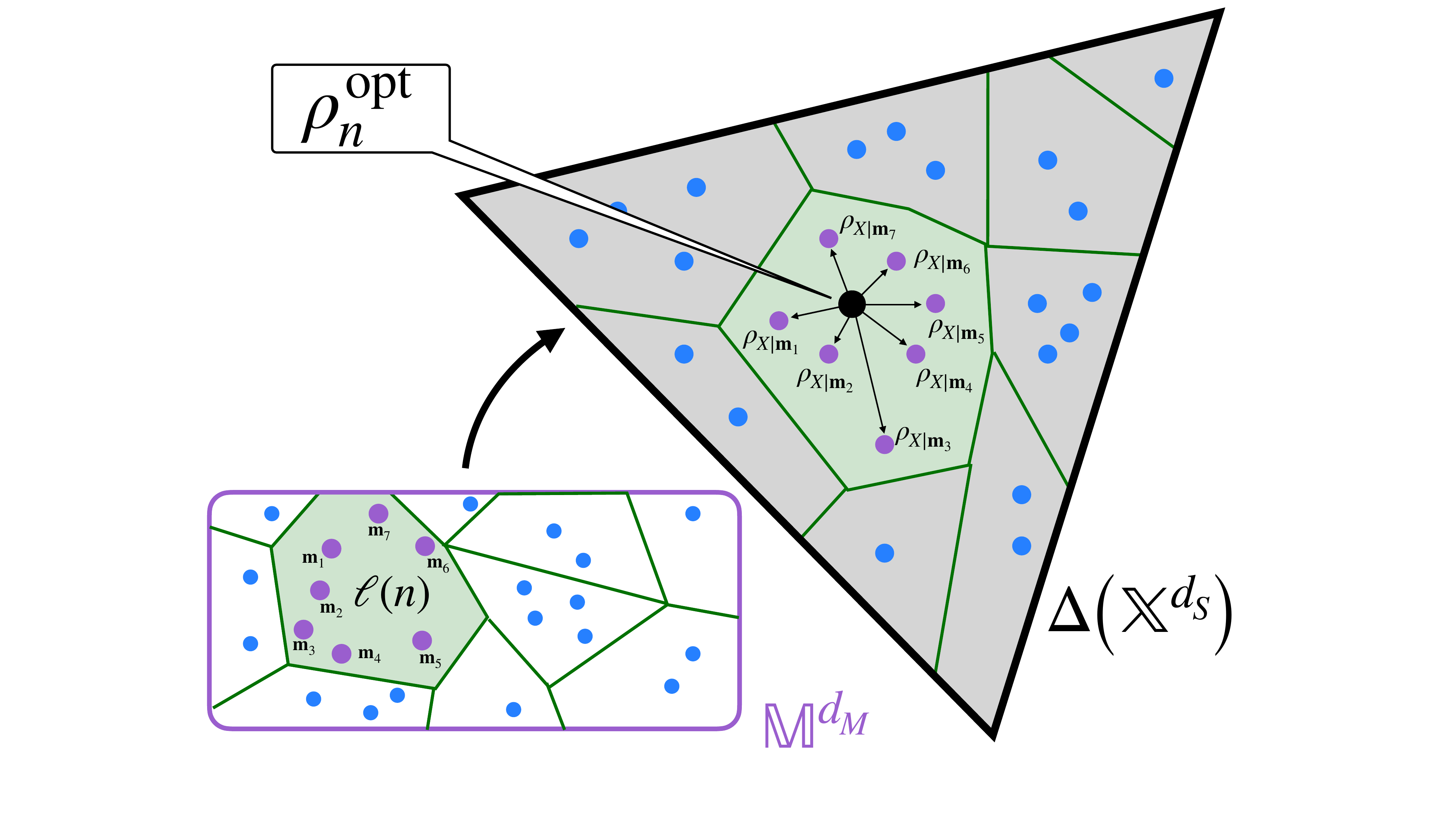}
 	\vglue -10mm
	\hglue -25mm	
	\hspace*{-1in}
	\caption{Schematic representation of social evolution and scaling of occupational diversity in cities by population (hypothesized).} 
\label{fig:5} 
\end{figure}

In the future, we intend to build a dataset of occupational diversity in deep time in order to better understand the relationship between social evolution and the total stock of information. We plan to exploit past and present census data and to deploy text mining methods on large textual datasets (Wikipedia, Google Books, Hathi Trust) to analyze the history of occupational specialization. 

Our hope is to be able to exploit this dataset to ask:

1) What factors account for increases in occupational diversity? Is it possible to distinguish the role played by market institutions, coordination costs, and the stock of information? 

2) To what extent is there cross-cultural structure in the changing composition of occupational specialties over time? Could Gutman scaling be used to identify necessary or probable sequences of occupations in the evolution of complexity (in the manner of~\cite{peregrine_universal_2004})? In other words, are societies universally likely to exhibit certain occupations in certain temporal sequences or other configurations?

3) Are there ways to exploit big textual datasets to understand the algorithmic complexity of specific occupations as well as the changing complexity of an occupation over time?

4) Is it possible to associate occupations with metafunctional categories such as production (e.g. metallurgist), consumption (e.g. entertainer), and information exchange (e.g. journalist), and to identify structures of change over time?

5) Is it possible to understand empirically how specific occupations interact in networks akin to the way that computers communicate (see following sections)?

6) Can changing occupational structures cast light on the macroevolutionary dynamics of human societies, revealing the relative and changing importance of processes such as convergent selection, conserved traits, horizontal transmission, contingency, the evolution of evolvability, etc. (as in~\cite{enquist_modelling_2011})?

7) Using time series or cross-sectional analysis is it possible to understand the ``bang per bit" (in the sense of~\cite{kolchinsky_semantic_2018}) of various occupations or algorithms or combinations thereof? In the long run, what kinds of increases in the stock of information have yielded the greatest changes in metabolism or complexity? Does the changing pace of change help illuminate the deep structure of social evolution? 

We believe occupational specialization is a suitable first proxy because crude historical data are conceivably attainable for deep time and because occupations plausibly reflect algorithms. However, we hope a computational approach to social evolution could inspire complementary efforts. It would be valuable to understand qualitatively and quantitatively how information storage and processing technologies have changed over time; how deep changes in the structure of science and engineering alter the ``bang per bit" of discovery; and how societies implicitly use algorithms to solve problems and how institutions shape social computation.

\section{A computational model of societies}
\label{sec:MCM_definition}

\begin{figure}
  	\includegraphics[width=1\linewidth]{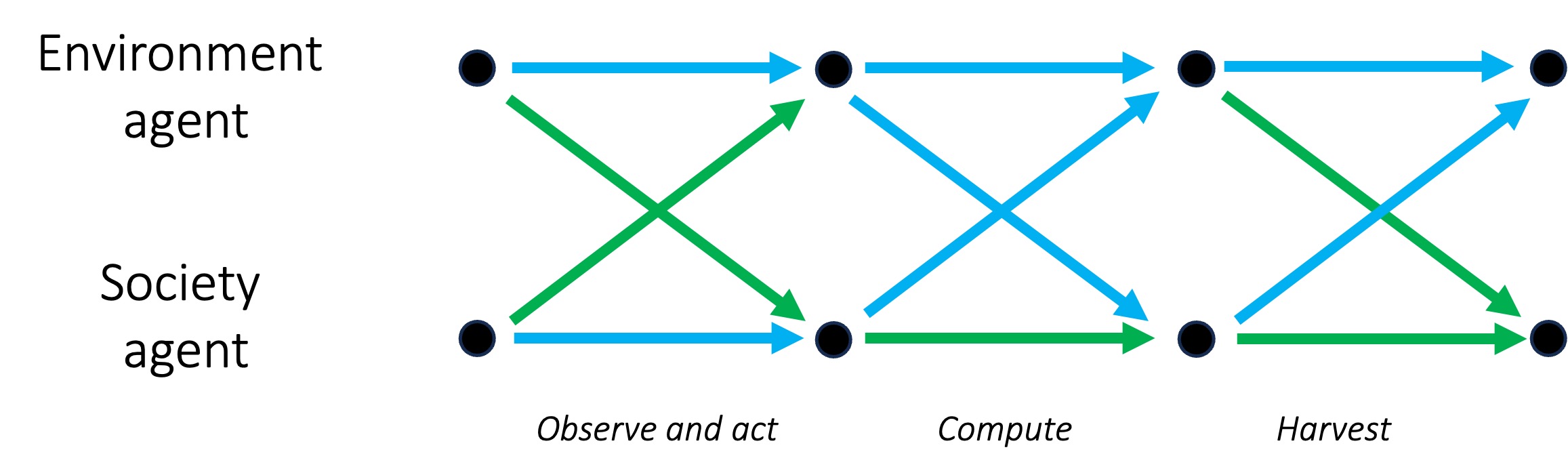}
 	\noindent \caption{Bayes net model of a single iteration of human social computer interacting
with an environment computer. Green arrows refer to actions actually taken by the Society Agent, 
as described informally in the captions. } 
\label{fig:6} 
 \end{figure}

Changes (positive or negative)
in the ability of a society to extract free energy from its environment (``harvest free energy'') affect its abilities to run computations, 
which in turn affects how much free energy it can harvest, which in turn affects its abilities to run computations, and so on.
(See the Bayes net illustrated in \cref{fig:6}.)
Empirical data show that this ``compute-extract'' loop is one of the primary drivers of the
long-term dynamics of human social systems~\cite{shin2020scale,morris2010}. However, at present
we have no \textit{theoretical} understanding of this loop. 
Our ultimate goal --- far beyond the scope of this preliminary paper --- is to start to fill in this gap.  

It is important to distinguish our goal  from the goals of already well-established bodies of
work that also involve both CS theory and the social sciences. Our goal is to
build a detailed model of the \textit{dynamics} of a social system as it makes its decision. We wish
to consider such models using the tools of computer science theory in general, and computational
complexity in particular.
In contrast, work in algorithmic game theory~\cite{niro01,roughgarden2010algorithmic}
or on the computational complexity of computing Nash equilibria~\cite{daskalakis2009complexity,papadimitriou2014algorithms}
focuses exclusively on the best possible performance (typically on the worst-case problem)
of a Turing machine at computing some abstraction of the end result of an infinite number
of steps of a real-world social system computer.
Our goal is to use CS theory to model the dynamics of a changing social
system, whereas  those other bodies of work instead use CS theory to derive results concerning the ``static''
properties of social systems, i.e., properties of their state at a single moment in time. 

Our premise is that a potentially powerful way to investigate the compute-extract loop
is to model both the human society and the  environment as computers, 
in the full CS theory sense of the term, that are iteratively interacting with each other. 
That might allow us couch the analysis  in terms of the
relationship between the attributes of two computational machines. It might also allow us to build upon 
the extremely rich body of theorems in CS theory. In the sequel, we will refer to each of those two computers
as an ``agent'', where as needed, we clarify whether we are discussing just the
Society agent or the Environment agent.

We draw inspiration for starting to pursue our goal from the field
of artificial life. One of the central  functions of living things is to store and transmit information, 
and to perform computation with that information, all the better to extract free energy from its environment. 
This conceptualization of biology led to the development of artificial life, first in theoretical work~\cite{schrodinger2012life,sipper1998fifty},\footnote{It is sobering
to appreciate that the artificial life community lauds Von Neumann for the creation of his ``replicator'', without any realization that
this work of Von Neumann's was
simply a restatement of Kleene's second recursion theorem of about a decade earlier. (Indeed, Kleene's  recursion theorem is
the foundation of the idea of a ``computer virus''~\cite{sipser1996introduction}.) Indeed, even though Kleene was a renowned logician,
Von Neumann failed to give him any credit.}
and then starting about half a century ago, in computational simulations~\cite{aguilar2014past,langton1997artificial}. 
This has led to great insights into the fundamental nature of all living systems. Our goal is to try to kickstart a similar
trajectory for archaeology and deep history, simply starting about 75 years after artificial life paved the way.

A central realization underpinning our approach is that there is major difference between \textit{communication} and \textit{computation},
even though social systems engage in both. Communication is all about trying to ferry bits from point $A$ to point $B$ with
as little distortion as possible. It is about information \textit{transmission}, and
involves error-correcting codes, Shannon information, mutual information, channel capacity,
and the other tools of the field of information theory. In contrast, computation involves information \textit{transformation},
and is all about \textit{changing} information, synthesizing different streams of information, etc. Rather than information
theory, the mathematical field that is central to understanding computation is CS theory. While communication and
computation are clearly closely related, and occur together in many natural systems (in particular in all 
social systems), they are fundamentally different.  Unfortunately, this is completely unappreciated
in the social science literature. Indeed, many people use the term ``computation'' and then
proceed to invoke information theory. We view our introduction of some of the concepts of CS theory
to the social science community as one of our most important (albeit trivial) contributions.

In the sections following this one we introduce a bare-bones model that we feel can serve as a seed for discussing
more fully featured models. The hope is that this model might be simple enough to allow some formal, mathematical
analysis, while at the same time have it be easy to enrich the model by adding insights from data sets and from domain expertise. 

%



The key feature of the computational process that we use is that each agent comprises a set
of multiple interacting subsystems 
that collectively perform the computation within a single
iteration. In particular, a human Society agent comprises (the members of) multiple interacting occupations, 
which collectively perform the computation the Society does to transform observations of the state of the
Environment into actions back on that environment.

The next subsection briefly describes some of the scenarios considered in CS theory that were
direct inspirations for our model for two agents that interact through
an iterative process. In the following subsection we introduce our formal model of the dynamics of
an agent within any single such iteration. The last subsection in this section then presents a set of
many examples of our formal model that occur in the natural world.

In the section after this one, we describe the
the way that two agents jointly interact, in the fully general case. We also describe the physical
meaning of the variables in this framework, paying special attention to how the framework
captures features of the computation process of an agent. In the following
section we describe several ways to tailor our framework concerning two interacting agents to 
the special case of one agent representing Society and one the Environment.

\subsection{Inspirations for our model of agent computation}
\label{sec:motivation}

The central feature of the model we introduce is that each agent (both the Society agent and the Environment agent)
comprises a set of multiple \textit{machines}. As described below, in this paper we view each machine in the Society agent
as a separate occupation, which could easily be reconstructed to include technologies as distinct machines. (However, we believe the model extends far more generally,
e.g., to usefully capture interactions among individuals, or among corporations, or among nations, rather than
only among occupations and technologies.) We then have the ``society computes'' part of the interaction between
the agents shown in \cref{fig:6} consist of some fixed number of ``timesteps'', in which all the machines in
the Society agent can communicate with one another while also performing computations in each timestep based on
what they currently know. In our model, that iterated sequence of interwoven communication  and computation is how the
agent as a whole computes. Accordingly, we call such an agent an instance of a ``multiple communicating machines'' (MCMs) model of computation.

There were several bodies of CS theory
that were inspirations for the MCMs model. Perhaps the primary one is work in CS theory on communication complexity~\cite{kushilevitz1997communication,rao2020communication,arpe2003one,babai2003communication,fischer2016public}. 
In standard communication complexity, there are $k$ machines that only send single bits to one another at each timestep, and they
must all end up calculating the desired function, $f(x_1, \ldots, x_k)$~\cite{rao2020communication}. 
%
There are several ways that the field of communication complexity has been extended to deal with more than two interacting
machines though. One, called ``simultaneous messages'' (SM), assumes that 
(like in epistemic logic) each machine $i$ can see $x_{-i}$, the joint input of all the 
other machines in the society (e.g., all the other bits being output by the environment).
They cannot see \textit{their own} input~\cite{babai2003communication}.
Instead, the work on SM 
has every machine
send a message to a \textit{referee}, who sees none of the inputs directly, and that referee makes the evaluation
of $f$. 
In our case, that sort of makes sense as a model of a hierarchy. 
%
In addition, there has more recently been work that is a ``number-in-hand'' variant of
this original ``number-on-forehead'' model. This new work is called ``private simultaneous messages'' (PSM), and has
every machine $i$ receive \textit{only} its own input $x_i$~\cite{ball2022note,fischer2016public}.

Communication complexity focuses on the minimal number of communication timesteps needed
to jointly calculate a function. We're actually interested in something different; how well one can do (e.g., 
the probability of success) at calculating the function if one has a fixed 
amount of allowed number of timesteps. In addition, we are not interested in having every machine
in the society agent calculate the exact same function.

Any modern computer continually runs multiple computations simultaneously, i.e., every modern computer is in fact a parallel computer. 
In this, they are exactly analogous to what happens in human social systems. Moreover, any such computation will often ``fork'' 
multiple new computations while it is still running, Any such fork increases the number of computations running 
simultaneously. This too is exactly analogous to what happens in human social systems.

In fact, how best to fork computations (or finish computations that are still running) is a central issue in the field of distributed computation.
An important special case of distributed computation (albeit one without forking of new computations, etc.)
is circuit complexity theory~\cite{arora2009computational,sipser1996introduction,jukna2012boolean}, which
is closely related to our problem.
Perhaps the easiest way to understand what this body of work is about is to consider Boolean functions.
A ``Boolean function'' $f$ over $n$ bits maps any possible sequence of $n$
``input'' bits to a single ``output'' bit. 
A ``Boolean circuit'' is a way to implement such a Boolean function.

To do this, a Boolean circuit arranges ``gates'' chosen from some fixed set of a few types of logical gate, 
like the logical gates AND, OR and NOT, and feeds those into
one another, in an acyclic fashion, to implement the Boolean function. 
Concretely, a ``Boolean circuit'' is a set of such
gates that are fed into one another, iteratively, through successive
layers. The first layer
is the sequence of $n$ bits that are the inputs to $f$, and the output bit
is given by the output value of the (single) gate in the last layer of the circuit.

There are many measures of the circuit complexity of any given function $f$. Examples
include the minimal number of gates of any circuit that can implement that function
for all possible input bit strings, and the minimal number of layers of any circuit 
that implements that function for all possible input bit strings.

Other work in circuit complexity theory considers circuits where the individual gates perform far more (or less!) 
rich information processing than simple Boolean primitives like AND, OR and NOT. The MCM setup
can be seen as such a circuit, where the individual gates are finite state automata, where each gate at
each layer has a specially identified parent gate in the previous layer, and sets its initial state to be the
ending state of its parent gate before processing all the inputs it receives from the other gates in that previous layer.
Finally, it is worth pointing out the work in~\cite{haastad2017average}, which considers circuits that are
provided with a particular computational task (related to ``Sipser functions''). For that task, if we measure
performance by considering the average case over all possible input strings, it turns out that reducing
the reducing the number of layers in the circuit by $1$ can drastically decrease performance. 



MCMs can also be viewed as a special type of a multi-agent system~\cite{romanowska2021agent}. 
In addition, the MCMs model can be viewed
as a special case of a collective~\cite{wola02,tuwo00,tuwo02,wosi01}, where rather than
an (exact) potential game~\cite{mosh96,candogan2011flows}, we have a team game~\cite{mara72}, and
the precise form of the communication among the members of the collective must obey multiple constraints.

In addition, cellular automata (CA~\cite{wolfram1984cellular}) can be viewed as a special case
of MCMs. This is done by
identifying each cell in a CA as a separate machine in an associated MCM. In general, an MCM has 
a far richer graph of inter-cellular direct communication, and a larger state space of each cell, than any CA.
In fact, so impoverished is the computational ability of cell in a CA, that information storage and processing
is typically identified only with inter-cellular dynamics, not with behavior occurring inside any individual cells~\cite{lizier2014framework}.
This is in stark contrast with the case of MCMs (and human societies).

A model that is more directly related to MCMs compared to CAs is known as the ``CONGEST model''
of distributed computation~\cite{hirvonen2021distributed}.  
The CONGEST model can be viewed as a type of MCM in which every machine can send $\log(n)$
bits of information to each of its neighbors in the message graph, in each timestep, where
(for example) $n$ might be the number of machines in the overall system. However, in
contrast to MCMs, in typical CONGEST models there are no restrictions imposed
on the computational power of the individual machines. Also,
the overall system's performance (its ``complexity'' to use the CS theoretic term)
is measured by the number of timesteps the overall system requires to solve the provided problem
perfectly. In contrast, as elaborated below, with MCMs performance is measured by the quality
of the overall system's response to a provided problem after a fixed number of timesteps.\footnote{In some ways 
this performance measure of MCMs is similar to what is called ``approximation complexity'' in CS theory~\cite{arora2009computational}.
See also~\cite{haastad2017average}}.)
Finally, the analyses of the CONGEST model usually consider worst-case
over inputs to the overall system, whereas with MCMs we are almost always
most concerned with the average performance over the possible inputs.\footnote{In some ways this feature of how we measure performance with
MCMs is similar to what is called ``average-case complexity'' in CS theory~\cite{arora2009computational}.}

A more direct inspiration behind the MCMs model than CONGEST was 
the idea of formalizing the computation done by 
a human society during each iteration as a computation by a Turing machine 
(TM~\cite{sipser1996introduction,arora2009computational,livi08}. Each iteration would
start with a new input provided to that TM by the environment, and that TM would then evolve in the usual
way through a succession of timesteps. After it completes the next iteration
would begin.  Particularly compelling is to consider a TM with multiple work tapes,
where each tape models a different occupation or technology in a society.

One difficulty with this approach is 
that TMs \textit{halt}. That means that there is no way for a society modeled with a TM to have memory
from one iteration to the next.
A natural way around this shortcoming is to suppose that the input to a society TM at the beginning of each iteration
includes its own state at the end of the previous iteration, in addition to getting information about the
state of the environment. However, unless that ``state at the end of the previous iteration'' is extended
to include the ending states of the work tape, there would be loss of memory of those ending states
from one iteration to the next. So
for example each occupation would forget what it had previously been thinking.
This problem can be circumvented if we instead model society  
as a monotone prefix TM (MTM), i.e., a TM that never halts, iterating forever,
where we stipulate that each iteration lasts for some fixed number $\tau$ of timesteps 
before the next iteration begins --- and with it arrives the next input from the environment.\footnote{Note
that imposing such a value $\tau$  is similar is similar to how an upper bound
on the number of allowed timesteps is imposed in the definitions of the usual computational complexity classes, e.g., $\bold{P}$ or $\bold{NP}$.}

However, even once we replace TMs with MTMs that run for exactly $\tau$ timesteps in each iteration, 
there are still two other substantial restrictions that we need to address if
we want to use our formalism to model the computation by a human society.
The first is that there are restrictions in how much the heads on each work tape can move along that tape in
a single timestep in an MTM, whereas there is no analogous restriction on how much the state of
an occupation or technology can change in a single timestep.
The second restriction is there are no direct ways in an MTM
model to allow different work tapes to talk directly with one another.




The model of the society agent that we investigate in this paper is a fully formal version of an MTM
that runs for exactly $\tau$ timesteps in each iteration, with the two restrictions mentioned just above
removed. We call this model a ``multiple communicating machines'' (MCM) model, using the term ``machine''
rather than ``work tape'', since the machines do not have restrictions on how they can communicate
with one another. 

Another important aspect of our investigation is to determine how a society reacts to information
from its environment to best act upon that environment. Our modeling of this aspect of human
social systems draws inspiration from the field of feed-back control. 
Feed-back controllers are artificial systems that react to stimuli in their environment. 
They underlie all of modern civilization’s functions; without them our modern society would immediately grind to a halt. 
(There are many feedback controllers in the reader’s home, in fact.) Moreover, aside from very simple 
feedback controllers, like thermostats, all feed-back controllers use embedded artificial computers to decide 
how to analyze their stimuli from their environment, and then act back on their environment. 
Again, the analogy between computers and human social systems is exact.

Ultimately, we would like our MCM model to also serve as our model of
the computation done by the environment. The motivation for using the same model for both systems is that it
might allow us to directly compare the computational
characteristics of the society agent and the environment agent. We might then
be able to analyze how their relative computational powers affects the ability of the
society agent to grow through time by interacting with the environment agent. In the sequel
though, we will focus on using MCMs to model human societies.

\subsection{Formal definition of Multiple Communicating Machines model}
\label{sec:single_agent}


In this subsection we first formally define MCMs, and then in the following subsection we provide a few
examples, illustrating the role the definitions will play in our investigation of
the interaction between human societies and the environment. 
Of course, in specific cases it might be worthwhile to investigate formal models where some of the following details
are simplified (or even removed), and / or some of the following details are elaborated in more detail.
We will refer to such models as \textbf{simplified} MCMs.

\begin{enumerate}
\item There are $N$ \textbf{machines} (or ``sub-agents'') in an MCM, indexed by values $v \in V$. 
The state space of machine $v$ is $X_v$, with elements $x_v$. 
The joint state of all the machines is written as $x \in X$.

We will abuse terminology and refer to ``cardinality of $X_v$'', or write ``$|X_v|$'', even though we mean different things
depending on the precise type of space $X_v$. For example, if $X_v$ is finite, then $|X_v|$ means the number of elements
of $X_v$, while if it is a Euclidean space $\R^n$ it means the dimension $n$, and if in fact the Cantor cardinality
of $X_v$ is at least $\aleph_2$, then $|X_v|$ means the Cantor cardinality of that set.

\item There is a directed graph $G$ with a set of $N$ nodes, $V$, each of which is identified with one of the
machines. An edge from node $v$ to node $v'$ is written as $(v, v')$.
The parents of node $v$ are written as $\pa(v)$, and the 
children are written as $\ch(v)$. $G$ is called the \textbf{message graph} for reasons that will become clear. 
\item There is a set of possible 
\textbf{external inputs}, $\EEE$, with elements $e$, which for simplicity is the same for all machines $v$.
\item There is a 
set of possible \textbf{(machine) messages}, $M$, with elements $m$, a set which  for simplicity is the same for all machines $v$.
\item As shorthand, for each machine $v$, a vector of messages $m_{v'}$ indexed by machines $v' \in \pa(v)$ 
is written as $in^v \in IN_v$. (Intuitively, during each timestep, machine $v$ will receive a message vector $in^v$ from
all the machines who are the parents of $v$ in the message graph.)
\item There is a machine-indexed \textbf{update} function
\eq{
f_v : X_v \times \EEE \times IN_v \rightarrow X_v \times M_v
\label{eq:update_func}
}

The update function $f_v$ has several arguments, which we consider in turn. 

The first  argument of the update function is just the machine's own state.

The second argument will be identified with an external input $e$, which concerns
the other agent (the one not containing the machine $v$), and which machine $v$ sees during each timestep in an iteration.
Note that this external input to $v$ is an extra input to $v$, in addition to the messages from
some of the other machines in the agent containing machine $v$. As described below, within any given iteration of the two-agent Bayes net,
the external input to agent $v$ will not change from one timestep to the next.

The last argument  is the vector of all the messages that $v$ receives from other machines in $v$'s agent. 
In general, these will change from one timestep in an iteration to the next.

The first output of the update function of machine $v$ is a state of machine $v$. 

The second output of the update function of machine $v$ is a message that machine $v$ sends to some other
machines (as specified in the message graph).

For simplicity we assume that each $f_v$ is a total (computable) function, guaranteed to finish between one timestep
and the next, for any triple of arguments to that function. 


\item To perform its computation,
the MCM runs for a total of $\tau$ \textbf{timesteps}. In each timestep $t$, all machine in the MCM
runs their associated update functions simultaneously. The state of each such machine $v$ represents
whether it has been activated (and potentially, how active it is, depending on our precise choice of $X_v$).
Machine $v$ uses the associated component of
the output of its update function to set its state in the next timestep, $t+1$. The other component
of the output of $v$'s update function is a single
message, $m_v$. This message forms the associated input argument in the next timestep to every machine $v' \in {\rm{ch}}(v)$. 
\end{enumerate}

Note that for simplicity, we suppose that the messages that each machine $v$ sends to all the machines in $\ch(v)$ each time
it runs its update function (i.e., in each timestep) are
identical, not varying among the receiving machines in $\ch(v)$. 
On the other hand, this single identical message that $v$ sends can change from one \textit{timestep}
to the next. And of course, it will vary among the machines $v$ in general.

\subsection{Examples of MCMs}

A concrete, physical example of a naturally occurring system that we can model as an MCM 
is a genetic regulatory network (GRN). In this example, each machine in the MCM model of the system is 
identified with one of the genes in the network. The messages in the MCM would be identified with transcription factors  the genes send 
to one another. So for example, if transcription factors are transmitted via diffusion, then the directed edges
in the message graph would be defined as the set of all pairs of genes $(v_1, v_2)$ where $v_2$ responds to the transcription factor 
released by $v_1$. (Note that in general this will be a proper subset of the edges in a fully connected graph
among the same set of machines, since not all genes respond to all other genes.)

The external input $e$ could then be some vector of counts of certain biomolecules from outside of the cell.
In real-world examples of GRNs, often there is no natural way to specify some
fixed number of timesteps,  $\tau$. This illustrates how in a real-world scenario, we might want to consider
a simplified MCM rather than a full one (cf. \cref{sec:single_agent}).

Another example of a system in the natural world that can be modeled with an MCM is a hunter-gatherer band.
In this example, each machine in the MCM model of the system is identified with an individual person in the band.
The messages in the MCM would be identified with things that the individuals say to one another (or in some
other way signal to one another). The external input $e$  could then be information that the members of the band
collectively gather about their environment. In contrast to the example involving GRNs,
quite often there will be a natural choice of $\tau$ in the case of a hunter gatherer band, e.g., if the band is hunting and must complete
the hunt before the sun goes down.

A related example --- central to this paper --- is a (complex) human society. There are many different versions
of this example, depending on what society-based informational systems we identify with the separate
machines. The example of such a process that we focus on in this paper is a human occupation.
There are many other processes that we include in our model as well though. For example, 
we could identify various technologies with some of the machines --- in this scenario, the ``messages''
back and forth between individual occupations  and technologies would model the uses of that technology
by members of that occupation. Another important process is information storage systems, e.g., libraries. So for example
we could use messages exchanged between individual occupations and libraries as models of members of those
occupations storing and / or retrieving information from those libraries. 
Other obvious examples of individual machines in a model of a human society  include  
specific bureaucracies, specific corporations, etc.

We can also model an environment that a human society interacts with as an MCM. The machines in that MCM could be systems like the weather, plagues, plants that might be receptive to domestication as food crops, etc. 
(As tongue-in-cheek, crude intuition, it may make sense to
identify each machine in the environment MCM with a phenomenon in the natural world
that polytheistic religions ascribed a deity to.) The external input of the Society MCM would be the observation
it makes of the state of the Environment MCM. 

In this example of a society and environment, we might also
invoke the Church-Turing thesis~\cite{piccinini2010computation,piccinini2011physical} and require that $f_v$ be Turing-computable for all $v$.
Also on physical grounds, no matter what the computational power of $f_v$, it makes sense to  require that the time to
run it is very small on the timescale of the MCM's timesteps, for any input $f_v$ might receive. 

Yet another physical example is a spin glass (or more generally, Potts model, or
some such) undergoing Glauber dynamics. In this example we identify each machine in the 
MCM  with a separate spin, with the message graph reflecting the coupling terms in the Hamiltonian
of the overall MCM. The external input could then be a magnetic field.

Moving on to the digital world, note that MCMs can be viewed as instances of parallel computers,
which are ubiquitous in modern-day digital computers. To see this we simply identify each
machine in the MCM with a different processor in a parallel computer, with the messages among those
machines identified with inter-processor communications. (We could augment this with details about shared memory among the processors, 
rules for how access conflicts are managed, etc.) In the same vein, MCMs 
likely have connections with circuit complexity theory, which
is closely related to the theory of parallel computation~\cite{arora2009computational,sipser1996introduction}.


As another more abstract example of an MCM, suppose that we allow $\tau$ to be infinite,
and have $e$ be an arbitrarily long sequence of bits. Suppose as well that every machine $v$ has an $\N$-valued
counter $n_v$ in it (i.e., $n_v$ is a component of the vector $x_v$), and that $n_v$ increases by $1$ in every timestep
for every machine. Finally, assume that  for all machines $v$, the associated update function $f_v$ can only depend on bit number $n_v$ in $e$.
In this case, each machine in the MCM is a Mealy machine. So the full MCM
is a set of $N$ simultaneously communicating Mealy machines.



Finally, it's worth pointing out that all of the motivating examples given in \cref{sec:motivation} (which do not
occur in the real world) can also be formulated as
MCMs. To do this though the machines in the MCM must have countably infinite state spaces (uncountably infinite,
in the case of a CA) and in the case of communication complexity, the machines
would typically be allowed super-Turing computational power.

Taken together, these examples demonstrate that while there is a fair amount of mathematical structure
in the definition of an MCM, it can still model a very wide range of real world distributed computational
systems. In addition, little of that mathematical structure is superfluous, in that most of it plays an important role
in the examples; simplified MCMs are often too impoverished
to capture important features in these examples.

Nonetheless, we emphasize again that many of the \textit{other} features of these examples
would be mostly unaffected by associated simplifications of the MCM model. If some such features are
the primary interest of the modeler, then it makes sense to simplify the MCM model accordingly.


\section{Interacting MCMs}
\label{sec:interacting_agents}

In actual physical systems, the exact same mathematical structure describes two processes
which are often misunderstood to be intrinsically different. The ``observation'' of agent $B$ by
agent $A$ is an instance of the state of agent $B$ at time $t$ affecting the state of agent $A$ at some later time $t'$.
(Formally, changes to $B_t$ result in changes in the conditional distribution $P(A_{t'} | B_t)$.)
Similarly, 
an ``action'' by agent $B$ on agent $A$
is an instance of the state of agent $B$ at time  $t$ affecting the state of agent $A$ at some later time $t'$.
So it is an ``observation'' by agent $A$ of the earlier state of agent $B$.
We can illustrate this with the example of a human Society MCM interacting
with an Environment MCM. In this case, the external input of the Environment MCM would in part reflect
the action that the Society MCM takes on the Environment MCM. (Such actions can range from types of niche construction
like constructing dwellings in the environment, to sowing seeds in the environment for future harvest.)

In this section we present a general framework of an ``iteration'' of two interacting MCMs. At the beginning
of such an iteration each MCM observes the other one (or equivalently, each acts on the other one). They then each carry out an
independent computation, by repeating their respective update functions (as described in \cref{sec:MCM_definition}). 
This gives us a basic, pared-down model of interacting MCMs, with an observation
process that also models control and action of one MCM by another.
%
%

The following section, \cref{sec:closing_loop}, presents a model built on 
this basic framework of interacting MCMs, tailored to 
the special case of one agent representing a complex human Society and one representing the Environment.
This model adds extra structure that formalizes
how the joint state of those two MCMs at the end of an iteration
can determine the free energy harvest by the Society from resources in the Environment,
with the resultant modification of the parameters of the Society MCM.

See \cref{sec:enriching_interactions} for some other ways to extend this basic
model of interacting MCMs.

\subsection{The joint interaction between MCMs}

Each of our two MCMs is defined by an associated collection,
\eq{
\Bigl\{N, G, \tau, \{M_v\}, \EEE, \{f_v\} \Bigr\} 
\label{eq:computation parameters}
}
When we need to consider the two agents simultaneously, we will index these features accordingly,
e.g., writing $\tau^{A}$ for the value of $\tau$ for agent $A$.

Loosely speaking, an entire run of $\tau$ timesteps of each MCM (one MCM modeling the society and one
modeling the environment) is interpreted as the computation by that MCM
occurring in each iteration of the Bayes net
depicted in \cref{fig:6}. Those computations of the two agents are done in parallel. So for example, 
if $\tau^{A} = \tau^{B} = 1$, then the two agents interact with one another continually, i.e., 
interact with one another just as frequently as the individual machines
in each agent interact with one another. More generally though, $\tau^A$ and / or $\tau^B$ may exceed
$1$, in which case the associated agent(s) spend some timesteps computing by themselves, without 
interacting with the other agent, before they interact again with that other agent.

The external input to each machine $v$ in an agent 
is interpreted as the result of the observation of
the other agent that the machine $v$ makes at the beginning of such an iteration.
This is why the external input does not change from one timestep
to the next in the formal definition of an MCM in \cref{sec:single_agent}.
As an example, this external input to some machine $v$ in the Society agent
is the attributes of the environment observed by the 
members of the machine $v$ at the beginning of an iteration. 
Note that because of this feature of the
interactions, if the values $\tau^i$ of both agents $i$ are greater than $1$, then the agents are performing their
respective computations faster than their interactions with the other agent. 

\begin{figure}
  	\includegraphics[width=1\linewidth]{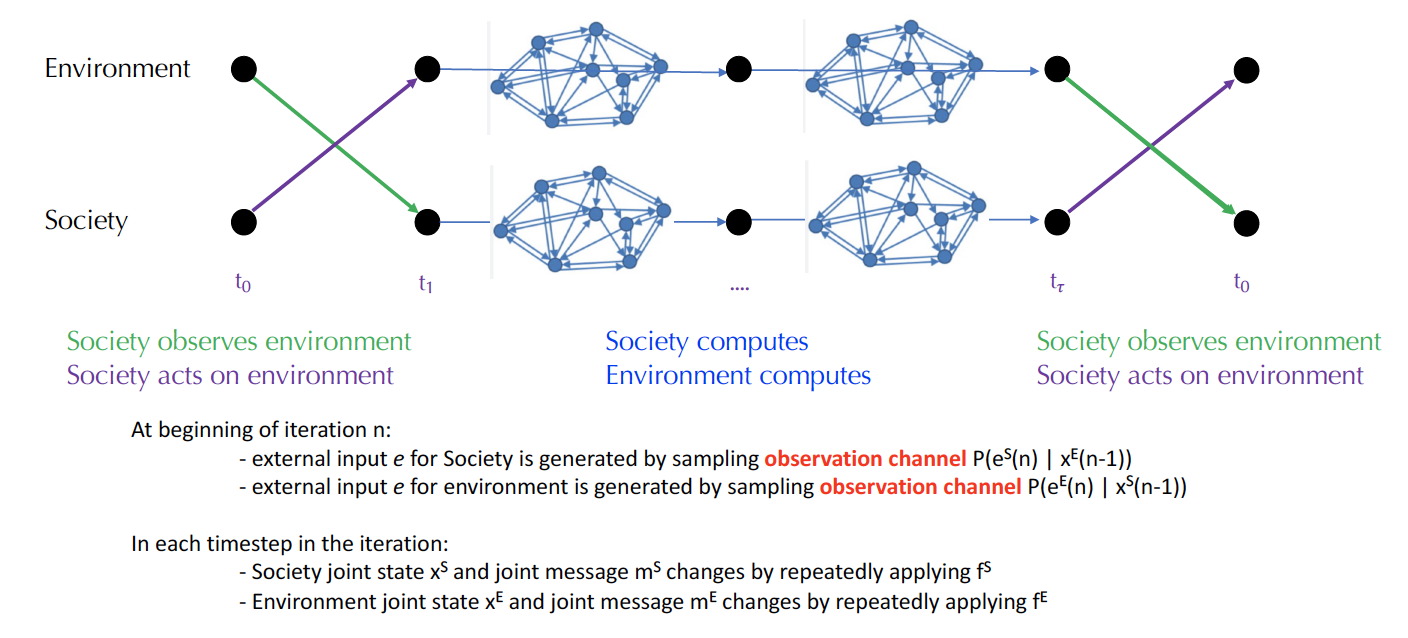}
 	\noindent \caption{Expanded view of the Bayes net model in \cref{fig:6} representing
an iteration of a human social computer interacting
with an environment computer. Green arrows refer to actions actually taken by the Society Agent,  and purple arrows
refer to those actually taken by the environment
as described informally in the captions. The messaging back and forth by the machines in each agent are informally
represented by the small directed graphs in the middle leg.} 
\label{fig:7} 
 \end{figure}

We now present the details of our model of the interaction of two MCMs :

\begin{enumerate}
\item An \textbf{iteration} of two interacting agents (MCMs), $A$ and $B$ consists of a pre-fixed set of 
$\tau^A$ timesteps for agent $A$ and $\tau^B$ timesteps for agent $B$. We adopt the convention that time
in each iteration starts at $0$, followed by timestep $1$,  timestep $2$ (if the iteration hasn't ended by then), etc. 

The Bayes net in \cref{fig:6} is a stylized depiction of an iteration, omitting the computation of the environment
agent, and omitting details of the computation by the society agent.

There are a countably infinite number of such iterations succeeding one another, forming a sequence. 
We adopt the convention that the joint state of the machines in the two
agents after the last timestep of iteration $t-1$ is the same as their joint state before the first timestep
in iteration $t$.

\item In the first timestep of iteration $t$, agent $A$ receives its external input from the final (timestep $\tau^B$)
joint state of the machines in  agent $B$ at the end of iteration $t-1$, 
according to some conditional probability distribution, 
\eq{
P(e^A(t) \,|\, x^B(t-1))
} 
This is the \textbf{observation channel}~\cite{coth91}
of the state of agent $B$ by agent $A$. This observation takes place in the first leg of \cref{fig:6}.
(See also \cref{fig:7}.)

Note that the external input to (all of the machines within) agent $A$ will be the same value
for all the timesteps in a given iteration $t$.  This contrasts with the messages, which may change at
each timestep within that iteration. This difference reflects the fact that messages occur as part of the single
computation that an agent $A$ does in response to the observations made by the machines $v$ in agent $A$
concerning the other agent, and these observations by the machines in $A$ are identified with the external inputs to 
those machines.

As an example, consider the case of a Society agent $A$ and Environment agent $B$.
In this situation,  $P(e^A(t) \,|\, x^B(t-1))$ models the observation that the members of the
occupations in the Society agent
are making throughout iteration $t$
of the state that the Environment had at the end of the previous iteration, $t-1$. 
(Note that notationally, in this conditional distribution the arguments of the $x$'s specifies the iteration, not the timestep
within the iteration.) In the case of a  environment,
agent $B$, $P(e^B(t) \,|\, x^A(t-1))$ models the effect throughout iteration $t$ on the
Environment of the joint action that the members of the Society agent took on
the Environment at the end of the preceding iteration, $t-1$.


\item After this first timestep, all machines in $A$ repeat their respective update functions
for the total of $\tau^A-1$ remaining timesteps, at which point iteration $t$ has ended.

\textit{Mutatis mutandi} for the external input to agent $B$ from the final state
of agent $A$ at the end of the previous iteration, and for its ensuing dynamics.

Note that if $\tau^A \ne \tau^B$, then we implicitly assume the two MCMs compute at different speeds, so that their iterations
both end at the same (wall clock) time, at which they take each take their actions / perform their observations.
For pedagogical simplicity though, in much of the discussion below we will take if $\tau^A \ne \tau^B := \tau$.



\item There is also a starting distribution over the joint state of all the machines in both agents, which gets
propagated in the usual way by the deterministic update functions of the two agents. For simplicity
we have not written it explicitly.

\item To simplify the exposition in \cref{sec:closing_loop} below, 
we will implicitly assume that the observation channel of each of the two agents is a symmetric noisy channel.
So each of those channels is parameterized by a single real number, setting the noise level. We write those
two noise levels as $\sigma^A$ and $\sigma^B$, giving the noise in the observation channel
of agent $A$ and of agent $B$, respectively.

\end{enumerate}

We will refer to the set of quantities concerning agent $A$ that are specified in \cref{eq:computation parameters},
together with the noise level $\sigma^A$, as the \textbf{computation parameters} of that agent.
As described below in \cref{sec:closing_loop}, all but one of
the computation parameters of the environment agent do not change from one iteration to the next.
However, as also described in \cref{sec:closing_loop}, the computation parameters of the society agent \textit{do} change between iterations,
as specified by the ``evolution function'' of that agent. 

\subsection{Physical significance of an MCM's computation parameters}

The computation parameters of an MCM 
determines its maximal computational power. In particular,
note that everything else being equal, the greater $N^{A}$ is, the greater the computational power of agent $A$,
(just like increasing the number of processing units in a parallel computer increases the power of that computer).
Note also that  the bigger the state spaces of the machines in an MCM is, the greater the memory space of each of those
machines, and so the greater the computational power of the MCM.

Similarly, the greater $|M^A|$ is, the more information the machines in the MCM
can send one another in every timestep. Accordingly, increasing $|M^A|$ increases the
computational power of that MCM.

Along the same lines, the maximal fan-out of the message graph of the machines in the society agent will
affect how much computation it can do in each iteration. As an example,
several of the ``complexity characteristics'' recorded
in the Seshat data set of the evolution of ancient societies~\cite{turchin_quantitative_2018}
were identified in~\cite{shin2020scale} as capturing the ``computational power'' of those societies.
These complexity characteristics included the sophistication of the writing system of the society,
the sophistication of its monetary system, and the sophistication of its transportation network.
The direct analogs of these in the interacting MCMs model
is the parameters $|M^A|$ and the fan-out of the message graph of agent $A$.

In addition, in the case of two interacting MCMs, the smaller $\sigma^{A}$ is, the more accurately agent $A$
can ascertain the starting state of agent $B$ at the beginning of an iteration.
Similarly, the smaller $\sigma^{B}$ is, the more accurately the agent $A$
can set a control action that it takes on agent $B$. 

In this regard, note that the action
that agent $A$ takes on agent $B$ at the beginning of iteration $t$ is based on the 
computation that agent $A$ completed in the \textit{previous} iteration, $t-1$.
That computation in turn reflects the observation
that agent $A$ made of the state of agent $B$ at the beginning of iteration $t-1$.
So loosely speaking, that agent $A$'s computation is how it produces its answer to the 
following question: ``If you observe
\{blah\} about the state of agent $B$ at a certain
time, what is the action you should take to optimally affect the state that agent $B$
will have by the time you finish answering this question?'' (The precise meaning of ``optimal'' 
in the context of a society-environment pair of co-evolving agents is discussed below.)

In real human societies, the population size is an extremely important variable. Indeed, the poorly chosen
term ``scalar stress'' (which is almost synonymous with population size) is ubiquitous in numerical data analysis in archaeology.
Indeed, variables involving
population size are rife in the 50 or so data fields recorded in the Seshat dataset~\cite{turchin_quantitative_2018,shin2020scale}.
Population size does not occur explicitly as a variable in the definition of an MCM. However, it does arise
implicitly, in several variables that \textit{are} in that definition. 

As an example, 
everything else being equal, if the population of ``laborers'' in a human society who are acting directly on the environment
increases, then the aggregate actions of those laborers on the environment is more precisely chosen,
with less variability. That can be interpreted to mean that raising the overall population 
of the laborers in a society results in shrinking $\sigma^{E}$.

Similarly, 
if the number of members of a specific occupation in a human society increases, and if those members are at least somewhat
distinguishable from one another, then that means that the joint state of the members of that occupation 
increases. In terms of the MCM formalism, that means a growth in the state space $X_v$ for
the machine $v$ that represents that occupation. Along the same lines, suppose the number of members of a specific
occupation increases, and they need not all send or receive messages
to the exact same set of other machines. In terms of the MCM model, this would mean
that increasing the population of that occupation causes a rise in the fan-out (out-degree) and fan-in (in-degree) of 
the node in the message graph that represents that occupation.

As a final comment, note that in the bare-bones version of the MCM framework presented above,
the external input $e$ of each machine in an agent does not change from one timestep to
the next during an iteration of the agent. This reflects a ``separation of timescales'' between the processes of computation
within each agent and the processes of each agent observing the other agent.

\section{Features of our co-evolving MCMs model and ways to enrich this model}

\cref{sec:MCM_definition,sec:interacting_agents} present a high-level framework for describing
interacting agents, with some discussion of the possible physical meaning of the variables
arising in that interaction. In many ways this is a minimal framework, 
providing no more than a way to capture the features discussed in~\cref{sec:motivation}. As described at
the end of \cref{sec:computational_approaches}, one of our goals in formulating this framework 
was that it not have too much structure to be amenable to mathematical analysis, and consequently be
``coarse-grained'' when considered as a model of an actual physical system.

However, one of our primary goals is to use this minimal framework to investigate 
the MET of the Anthropocene
(and more generally of human societies in the latter part of the Holocene). To do that we need to introduce extra mathematical
structure, specifying how the parameters and functions
defining the Society agent evolve in time, in response to the society's interactions with the Environment agent.  
As discussed at the end of \cref{sec:computational_approaches}, this extra structure results in a ``moderate-grained'' model
of the interactions between complex human societies and their environment.

There are very many details that seem unavoidable in such a moderate-grained model, unfortunately. 
To keep the discussion in the main text coherent, we present one such moderate-gained model
in \cref{sec:closing_loop}, rather than here in the main text. 
For simplicity and clarity, from now now on we no longer talk generically about two co-evolving agents $A, B$.
Instead we consider the specific pair of a Society and Environment co-evolving agent,
which are the focus of our study. We will distinguish the computation parameters and functions of those two agents
with the labels $S$ and $E$, respectively.

The underlying idea of our moderate-grained model
is that it takes energy to maintain high levels of
each of the computation parameters of the Society agent, i.e., there is an energy-cost function
associated with each computation parameter. In addition, at the beginning of each
iteration the Society MCM will ``harvest'' a certain amount of energy from the Environment MCM. 
We refer to this as the \textbf{gross free energy harvest rate} (GFER). The GFER must cover all
the thermodynamic costs of the society's computation (as well as many other thermodynamic
costs, e.g., maintaining homeostasis of the members of the society). 

The coupling between
the thermodynamic harvesting and the thermodynamic expenditures is achieved via the computational power of the society.
The GFER is determined by how well the Society can control the Environment MCM, acting
on the Environment in a way that causes it to have a desired state in the future. In turn, the 
maximal ability of the Society agent
to exert that control is determined by the best computation it can do, and so by the values of the computation parameters
of the Society MCM. In sum, how much computation the Society can perform and how well it can use the results
of its computation to control the Environment determines how much energy the Society can harvest from the Environment.
That energy harvest in turn determines how much computation the Society MCM can perform, and how well it can
can use the results of its computation to control the Environment.

One high-level point to keep in mind is that the MCMs model is a \textit{model}. The precise mapping of the
variables in the model and their dynamics to associated phenomena in the real world is not specified, and
may in fact vary from one use of the model to another. In particular, the ``energy harvest'' process at the end / beginning
of an iteration takes one timestep in the model. However, in the real world, it may take an extremely long time,
e.g., if a semi-static thermodynamic process is used to maximize efficiency. All such details are elided below.

There are numerous other salient features of the co-evolving MCMs model which are worth elaborating, in
addition to those that go into our ``moderate-grained'' model. 
There are also many features that we may want to introduce to extend this model. 
In the rest of this section we present both those features of the MCM model.

\subsection{Features of the interacting MCMs model}

There are several features of the interacting MCMs model that are worth emphasizing.

\begin{enumerate}
\item 
If $\tau^{S}$ is large while $\tau^{E}$ is small, then the Society does a lot of
computation between two successive times when it acts on the Environment --- and the Environment does \textit{not}
do much computation between those times. So in this situation, the Society agent is computing
what to do quickly on the timescale of the dynamics of the Environment.

\item The fact that our model allows the technologies to change from one iteration
to the next  might provide a way of incorporating Guttman scaling~\cite{forrester2009guttman,peregrine2018toward,peregrine2004universal}
into our analysis, i.e., of incorporating the idea
that some sets of new technologies can only be added in a particular sequence, with each
step in that sequence enabled by an increase in the GFER at the beginning of an iteration. One way to 
integrate this feature into the model is to ensure that the dynamic functions evolution operator only adds
machines with certain capabilities in such a Guttman sequence. 

\item 
One of our primary goals in formulating the interacting MCMs model is to use it to investigate the explosive escape from
Malthusian traps that has happened in human society in the last half century, starting in the West.
(Some authors refer specifically to the gap that opened up between the West and China 
as ``the Great Divergence''~\cite{pomeranz2000great}, whose dating is highly disputed~\cite{scheidel2019escape}.) In terms of our interacting MCMs model, that escape corresponds
to a super-linear rate of growth of GFER. 

Accordingly, to investigate that escape
we need the pair of the evolution function $\phi$ and harvest function $\psi$
to have the property that --- for a certain very restricted set of their input values
(i.e., a very restricted set of joint values of the GFER and the computation parameters of the society agent) --- 
the effect of the output of $\phi$ on the joint dynamics of the Society and Environment agents
causes yet higher GFER in the next iteration. In particular, it would be very interesting if
in the very next iteration, the GFER increases faster than population
size, so that the GFER per capita rises, before the population increases enough in the subsequent
iterations  without any more rise in the GFER so that the GFER per capita falls
back to its original value. This would be a formalization of escaping a Malthusian trap.

\item Ideally, we want there to even be situations, i.e., joint states of the two agents and associated GFER,
such that once the pairs of agents hits one of those joint states, $\phi$ together with $\psi$ causes 
a ``run-away'' dynamics, where the GFER starts increasing exponentially faster than the population size
from one iteration to the next for a sequence of multiple iterations.

Phrased differently,
we want our model, and in particular this pair of functions, $(\phi, \psi)$, to endogenously capture a phenomenon
whereby once a threshold is reached, one has a ``punctuated equilibrium'', of a sort, i.e., a discontinuous
leap to a new level of computational power. In our case, to model the current major
evolutionary transition, we want these leaps to cascade, i.e., we want each one quickly lead to a next one, etc., 
in a quickening avalanche of such leaps. 


\item 
Note that this property may already be built into our model. As the Society agent adds
new machines from one iteration to the next, the associated message graph may undergo phase transitions, from having only small,
disconnected components to having giant components, and eventually to being fully connected. This would
correspond to a ``jump'' in the computational power of the Society MCM.
Similarly, as $\tau^S$ increases and / or the number of machines in the Society MCM
increases and / or its message graph's fan-out grows, the computational power of the Society MCM increases. 

For a fixed Environment MCM, such increase in the computational power of the Society MCM might 
go through a ``phase transition'', suddenly allowing it to 
predict the future state of the Environment MCM far more accurately than before. Alternatively, reductions
in $\sigma^E$ might cause a jump in how well the Society MCM can control the future state of the Environment
MCM, thereby simplifying the computational difficulty of predicting the future state of the environment. 

In all of these cases, the jump in the ability of the Society might lead to a jump in the value
of the GFER that the Society agent acquires at the end of the iteration. (For example, one would expect such a jump
in the case of the mutual information harvest function, described in \cref{sec:evolving_parameters}.)
These jumps might then ``feed on themselves'',  resulting in an accelerating sequence of jumps in the value of the
GFER, and therefore in the energy usage per capita of the society.

%

\item Note that it is pretty much incontrovertible that 
societies don't really consider the costs associated with the future, only the benefits. Moreover,
they typically only look a very short distance into the future. So rather than than try to increase  
something like the expected future-discounted GFER (which depends on future costs), they
try to increase the immediate (next iteration) GFER. For us this might be an advantage, since
it presumably substantially simplifies the approach to setting the computation parameters
and dynamics functions and allocation distribution to be optimal.

\item More generally, independent of what human societies actual do or don't try to optimize, there might be some rich mathematical analysis 
even if we restrict attention to a single iteration of 
the co-evolving MCMs, without considering the evolution function of the society agent or the nature of the environment 
agent's dynamics across iteration boundaries.

In particular, if the environment were 
dynamically simple ---  that is, if it were simple when viewed as an MCM ---
then intuitively, one would expect that there would be no benefit to human society of increasing its computational power 
over multiple iterations. It would be interesting to either establish that this is the case, or show that it is not.

\end{enumerate}

\subsection{Extending the MCM model}
\label{sec:enriching_MCMs}

%
The definitions in \cref{sec:single_agent} only provides the most basic type of MCM.
Many of the features of this basic type of MCM are chosen for pedagogical simplicity, and
to open the possibility of mathematical analysis (e.g., of scaling behavior). 
This definition can be enriched
in many ways, to define a more complete (and therefore more complicated) type of MCM.

\begin{enumerate}
\item In the basic type of MCM, the messages that $v$ sends to all the machines in $\ch(v)$ each time
it runs its update function (i.e., in each timestep) are
identical, not varying from one receiving machine to the next. In the full model, that restriction is relaxed,
allowing each machine $v$ to send a different message to each of its child machines in the message graph.

\item Similarly, the message spaces of the machines in an MCM are identical in the basic type of
MCM, whereas they are allowed to differ from one another in the more complete definition of an MCM. 


\item In the basic type of MCM. the messages that any machine $v$ sends to all the machines in $\ch(v)$ each time
it runs its update function (i.e., in each timestep) are identical. This  means that
we do not allow a machine to ``tailor'' its messages to the recipient. A natural way to enrich that simple  MCM model
would be to allow such tailoring, i.e., allow $v$ to send different messages to the machines in 
$\ch(v)$ every time it runs $f_v$.

\item In the real world there will not be an update (deterministic) \textit{function},
 $f_v$, but rather an update \textit{conditional
distribution}. Again for pedagogical simplicity, the simple MCM model does not have this flexibility,
i.e., it presumes that all such distributions are delta functions. An obvious way to enrich
the basic MCM model so that it is more realistic is to allow the $f_v$ to be conditional distributions.

%

\item Another way to enrich the basic MCM model is to have a special machine that serves as a \textbf{ledger}. 
This would allow us to model social institutions like the library for Ashurbanipal,
or the clearing house of receipts in many ancient Levant societies --- or the internet for modern society. Going
further afield to consider different METs, if we model a single eukaryotic cell 
as an MCM (with the machines being organelles), the ledger could be the 
(epi)genome. 

In whatever context, a ledger machine's state space would be a tape 
(as in conventional TM theory), which may either be infinite, or finite (as in a genome)
or finite and growing with the number of iterations, as in a library. In addition, what could make the
ledger particularly useful in our model of a society agent is that we could choose a message
graph such that (almost) all non-ledger
machines would be able to send messages to and receive messages from the ledger machine.
The effect on this ledger machine of the messages it receives
from any other machine in a timestep is equivalent to that other machine
writing to a specific location on the ledger machine's tape, and / or sending 
requests to read the contents of a specific location in the ledger machine. 
The state of the ledger machine would change appropriately in response to write messages. 
The messages sent by this ledger machine would then be different for each of the other machines, given
by the ledger's response to the read request just made by those other machines (if any).

\item 
\label{item:7.2.6}
Similarly to a ledger, we could also have various machines that represent stores of material resources of various sorts.
We could then have a message to or from the machine enforce conservation of the associated good. For example, 
suppose a laborer machine $v$ mines a  single unit of some material resource from the environment at the beginning of an iteration.
In the MCM framework, that means that at the beginning of the iteration, one component $j$ of
the observation vector, $e(j)$ of the society MCM has the value $1$, specifying 
that single unit of that resource that was mined. In the first timestep, machine $v$
increases some associated component $i$ of their state vector, $x_v(i)$ by $1$. For them
to then send that unit of the resource to a storage machine $v'$ would mean that in the associated
timestep they send a message with the value $1$
to $v'$ and decrease $x_v(i)$ by $1$ at the same time. Then in the next timestep, the value of $x_{v'}$ increases by $1$. If
at some subsequent timestep some other machine $w$ needs to use that resource, a message can be sent from $v'$ to
$w$ accompanied by a drop of $1$ in the value of $x_{v'}$, and then the associated component of $x_w$ increases
by $1$ when they receive that message in the next timestep.

More generally, production functions (in the standard economics sense) could be modeled as resource machines that 
send messages directly one another. The update function of a resource machine, setting the new value of its state
based on the messages it just received and its previous state, would just be the associated
production function we want to model with that machine.

%

\item Yet another way to enrich the model is to restrict the possible form of the message graph
of an agent so that some nodes have far greater maximal fan-out than others.
With that extension, introducing a new communication technology in a society agent 
would simply amount to introducing a new machine whose fan-out is far greater than that
of the other machines in the agent.
\end{enumerate}

\subsection{Extending the model of interacting MCMs}
\label{sec:enriching_interactions}

The type of interaction between MCMs defined in
\cref{sec:interacting_agents} is an very pared-down, basic one. It can be enriched
in many ways, to get a more complete (and therefore complicated) model.

\begin{enumerate}
\item One natural way to enrich the basic co-evolving MCMs model 
sketched above  would be to have the noise in the two observation channels not 
be uniform and symmetric. For example, this would provide a way for the Society agent to
pay especially close attention to some of the variables in the environment. 

\item More generally, each machine $v$ in an agent could get a different external input from the timestep $0$ state
of the the other agent, produced with a different observational distribution over values of  $e_v$ conditioned on the
joint state of the machines in the other agent. That set of machine-indexed observational channels would provide a 
\textbf{observation graph} for the agent.

As an example, providing the Society agent with a observation graph would provide a way for some of the occupations in
the Society agent to pay particularly close attention to the values of some subset of the variables in the 
Environment agent, e.g., due to geographical proximity. It may even be
that some other occupations in the Society agent do not pay direct attention to
\textit{any} of the variables in the Environment. 

\item Similarly, recall that the observation channels of the Environment agent
are the same as the action channels of the Society agent. Accordingly, restricting which of the nodes
in the Society agent are roots in the Environment agent's observation graph would allow some of the occupations in the Society agent
to be more closely involved in acting directly on the Environment compared to the other occupations. 
%
This would be a way to capture the fact the members of different occupations observe (and affect)
different variables in the environment (due to physical location of the members of the occupation,
if nothing else).

\item Physical location can be quite important for how real human societies evolve from one timestep to the next, not just 
for how they can act on the environment at the end of an iteration. This raises the issue of how to
represent the spatial locations of the machines in an agent. One natural way to do it would
be by having the message graph of the Society MCM be a planar graph, with only
nodes that (represent machines that we want to model as) physically close to one another directly connected to one
another by that message graph. 

\item Another important point to note is that in the basic model of the interaction of the Society and Environment agents
in \cref{sec:interacting_agents}, there is a ``harvest function'' that specifies how the Society agent extracts free energy
from the Environment, leading to new values of the parameters of the Society agent. However, there
is no corresponding function that specifies how the Environment agent's state changes when free energy is 
extracted from it.  

A natural extension of the basic model would capture this by providing special ``resource storage'' machine in the 
Environment agent, whose
value specifies the amount of the resources whose harvesting provides free energy to the 
Society agent. There would then be a new
step, just before the beginning of any iteration but after the end of the previous one. In that step 
the value of the resource storage machine in the Environment is reduced by the GFER generated by
the harvest function of the Society agent (which in turn is based on the ending joint state of the two MCMs in the previous iteration).

\item Similarly, in a more elaborate model of the interaction of the Society and Environment agents, there might
be resources which are transferred from the Environment agent to the Society agent but do not directly change the
computation parameters or dynamics functions of the Society agent. These resources could be things
like minerals, petrochemicals, building materials, timber, etc. 
To model a transfer of these resources,
we would simply have machines in each agent that represent the total storage of the associated resource
in that agent. At the beginning of an iteration, there could be a transfer of such resources between (the storage machines
of those resources in) the two agents. Formally, this could be implemented without adding any more mathematical
structure to the model of interacting MCMs above, simply by carefully incorporating ``corresponding'' dependencies
in the observation graphs of the two MCMs, ensuring that the values in the associated machines in the MCMs changes
under those two observation graphs in such a way
that an increase in one results in a decrease in the other.

\end{enumerate}

\section{Discussion}
\label{sec:discussion}

The goal of this paper has been to identify the major gap that exists in the study of human social evolution. 
Namely, due to a series of contingencies in intellectual history, the study of social evolution has been 
disconnected from advances in computational complexity. It seems self-evident that there are advantages to bringing these two bodies of thought into conversation. It is much harder, of course, to bring this reunification about.

Here we have proposed a combined theoretical / empirical strategy to get at the issue. We think that our proxy, occupational specialization, has the potential to make concrete and visible the flow of information within a society and across evolutionary time. We have also proposed a formal, mathematical model of how a society computes and how its computational power changes over time. We have not, obviously, started the merger of our data and our model. That comes next, after the candid feedback of our august assembly, for which we express our advance gratitude.

There is at least one major body of work involving information theory 
to model the evolution of a species extracting resources from its physical environment.
This body of work involves ``Kelly gambling'' and ``bet-hedging'', and is discussed in~\cref{sec:kelly_and_beyond}.
It would be straightforward to incorporate that body of work into our framework,
but we do not consider it here, for reasons of space. 
(See also~\cite{guttenberg2008cascade} for a formal model that could maybe be usefully applied to investigate the METs.)

In contrast, we could not find any way that Ashby's  semi-formal ``requisite variety'' idea could
enrich our framework. In fact, Baez and Aaronson (in separate blog posts) argue strongly that Ashby's idea is either
tautological or wrong, depending on how one carefully defines it.


\section*{Acknowledgements}

DHW and KH would like to thank the Santa Fe Institute for support.

\appendix

\section{Appendix A: Integrating free energy harvest into the interacting MCMs framework}
\label{sec:closing_loop}

\subsection{Evolving the parameters of the MCM computation}
\label{sec:evolving_parameters}

In this subsection we specify one way that the parameters of the computations by the
two agents can evolve from one iteration to the next.

%

\begin{enumerate}
\item 
The joint state of the Environment and Society at the beginning and end of an iteration determines the value of
the 
GFER of that iteration, via a
\textbf{harvest function} for the Society agent, $\psi$. 

To ground thinking, note that in many ways a human society needs to act on the environment in
a way that causes the environment to have a predicted state in the future. In terms of interacting MCMs,
this means that by appropriate action on the Environment MCM at the beginning of an iteration, the Society MCM
is able to cause that Environment MCM to have a value at the \textit{end} of the iteration that matches a prediction for
that ending state that is made by the Society MCM, at the \textit{beginning} of the iteration. This can be formalized
simply as the mutual information between those two states, $I(X^{E}_{\tau^{E}}; X^{S}_{0})$.

There are many other choices for the GFER though. Some of them are actively grounded in stochastic thermodynamics,
and in fact are specifically designed to model the case of a physical system acting on its environment so as to
extract as much free energy from that environment as possible. Preliminary analysis
of a highly simplified such system, in which the agent has only a single machine and a single timestep, has proven surprisingly
fruitful~\cite{hartle2024work}. Extending this preliminary analysis is an active topic of research.

\item Suppose we are give a value of the GFER at the end of an iteration. We need to
allocate that GFER among the computation parameters defining the Society agent, to set the values of
those computation parameters in the next iteration. 

Intuitively, this allocation reflects two features of our framework. First, as mentioned, many of the
computation parameters of the Society agent have an associated thermodynamic cost, which
scales with the value of that computation parameter.
Second, for simplicity we do not allow the Society agent to somehow store unused free energy
from one iteration to the next. Accordingly, the GFER must be completely ``spent'' in each
iteration by paying the thermodynamic costs given by the precise values of the computation parameters
of the society agent. Together, these two features uniquely specify the values of all the computation parameters
at the start of the next iteration. 

We next present a discussion of the computation parameters of the Society agent whose values have associated
thermodynamic costs. Then we present a simple formalization of how to 
allocate the GFER among those computation parameters.
Crucially, the cost of each of these computation parameters is an invertible function of the value
of that computation parameter. As a result, the fraction of the GFER allocated to any given computation parameter
will set the value of that computation parameter for the next iteration.

\item The computation parameters of the Society agent that cause thermodynamic costs in each iteration are as follows:
\begin{enumerate}
\item 
There is a thermodynamic cost function that monotonically increases with increasing $\tau^{S}$,
since the larger $\tau^{S}$ is, the more total computation the Society agent does per iteration.
Overloading notation, write that function as $C_{\tau^{S}}$.

\item 

There is a thermodynamic cost function that monotonically increases with increasing  $|M^{S}|$,
since the larger $|M^{S}|$ is, the more costly it is for a machine in the Society agent to send a message which
requires it to choose among the $|M^{S}|$ possible value.
(E.g., in a human society, $|M^{S}|$ could be the
number of possible postal letters or other forms of communication that one machine
can send to another in each timestep.) 
Overloading notation, write that function as $C_{|M^{S}|}$.

\item  There is a thermodynamic cost that monotonically increases with increasing $N^{S}$, 
the number of machines in the society MCM, i.e., increasing
the number of occupations and technologies. (This cost arises from the need to perform
maintenance on the members of the occupation, i.e., due to the costs of maintaining
homeostasis.) Overloading notation, write that function as $C_{N^{S}}$.

\item There is a thermodynamic cost function that monotonically increases with increasing
the number of possible states of each machine in the society agent, $|X^{S}|$.
(That's because like the number of possible messages, increasing this computation parameter 
increasing the number of states each machine must choose among from one timestep
to the next.) Overloading notation, write that function as $C_{|X^{S}|}$.

\item There is a thermodynamic cost function
that monotonically {increases} with increasing $r^{S} : =1 / \sigma^{S}$,
i.e., with decreasing the noise in the observation channel through which the society MCM observes the environment MCM.
Overloading notation, write that function as $C_{r^{S}}$.

\item There is a thermodynamic cost function
that monotonically {increases} with increasing  $r^{E} : =1 / \sigma^{E}$
the noise in the observation channel through which the \textit{environment} MCM observes the 
\textit{society} MCM. That's because as discussed below, for MCM $A$ to observe MCM $B$ is
formally identical to having MCM $B$ control MCM $A$, and so decreasing this level of noise
is formally identical to decreasing the amount of noise with which the Society agent acts on
on the Environment agent at the beginning of an iteration. (This is the only computation parameter of
the Environment agent that will change from one iteration to the next.)
Overloading notation, write that function as $C_{r^{E}}$.

\item 
There is a thermodynamic cost function that monotonically increases with increasing maximal
fan-out (i.e., out-degree) of the message graph of the society agent, $D(G^{S})$. (That's because this 
feature of the message graph
is the maximal number of recipients of a message by a machine in the society agent during a timestep.)
Overloading notation, write that function as $C_{D(G^{S})}$.

\end{enumerate}
In some circumstances we can  set these cost functions based on real-world data or domain expertise. Those
that can't be set that way can be set in any mathematically reasonable
way. For example, we can simply take them to be the identity function 
(i.e., take $C_i(z) = z$ for all $z$ for such computation parameters $i$),
Another alternative would be to set $C_i$ to equal some simple concave function,
if with think there are economies of scale association with that computation parameter, 
e.g., take $C_i(z) = \sqrt{z}$ for such computation parameters $i$. Yet another possibility would be
to set $C_i$ to be a logit  function, if we think there is a maximal value of the thermodynamic cost for some computation parameters.

\item The \textbf{(computation 
parameter) evolution function} $\phi(GFER)$ specifies the values of the computation parameters in the next iteration,
based on the GFER at the end of the current iteration.
An example of such evolution function, chosen for simplicity, is the following:

First, define an \textbf{allocation distribution} $\rho$ over the possible joint values of the
computation parameters.
So $\rho$ specifies a fraction for each computation parameter, where the sum of the fractions equals $1$.
As shorthand define $R_t(i)$ to be the product of the GFER at the end of iteration $t$ and $\rho(i)$, the fraction
of that GFER that is allocated to parameter $i$. Then 
\eq{
\phi_\rho(GFER_t) = C^{-1}_i(R_t(i))]
}


This reduces the specification of how the computation parameters evolve from one iteration to the next to the specification
of how the allocation distribution is set for each iteration. How to set that distribution for the Society 
agent is discussed in the next subsection. 

We will not consider such an allocation distribution for the Environment agent however. That is because as discussed below,
we will take all of the Environment agent's computation parameters to be unchanging from one iteration to the next.
(The only possible exception is $\sigma^{E}$, whose evolution is actually set by the evolution function of
the Society agent.)

\end{enumerate}

\subsection{Evolving the functions of the MCM computation}

The preceding subsection only specified how to set the computation parameters, given a harvest function and an allocation distribution.
That still leaves open the question of how to set that allocation distribution. It also leaves open the issue of how
to set the associated
functions of the two agents which involve the computation parameters. Only once we have specified
these details will we have fully specified the co-evolving dynamics of the two MCMs.

\begin{enumerate}
\item To begin, label the iterations as $\{0, 1, 2, \ldots\}$.
For the evolution function defined above, the values of the computation parameters in any iteration $t+1$ where $t \ge 1$
are uniquely fixed by the value of the GFER at the end of iteration $t$.\footnote{Note that this still requires us to set the values of those
computation parameters at the beginning of the first iteration. Similarly, we need to set the joint distribution over the
states of the Society and Environment MCMs at the beginning of the first iteration. Such details are not discussed in this paper.}
In turn, the GFER at the end of iteration $t$ is set by the joint distribution over the states of the society
agent and the environment agent at that time. 
Finally, that joint distribution is itself ultimately set
by the combination of the values of two quantities at the beginning of the first iteration.

The first of these two quantities is the joint distribution of the states of the machines in the two agents at the
beginning of that iteration. The second is the functions which (together with the Society agent's initial computation parameters) 
specify the computation within each iteration. Those \textbf{dynamics functions} for the Society agent are 
\eq{
\{f_v^{S}\}, G^{S}, \rho
}
Similarly, the dynamics functions of the Environment agent are 
\eq{
\{f_v^{E}\}, G^{E}
}

\item We can take the dynamics functions of the Environment to be invariant from one iteration to the 
next, along with the computation parameters of the Environment. 
If we do this, then loosely speaking, Society can act on the Environment by changing the values of some of the Environment's
variables, which in turn affects the dynamics of the environment, but Society cannot directly
change that dynamic process of the environment itself.

\item Specifying those (time-invariant) computation parameters and dynamics functions of the Environment agent
is a perfect place to insert domain knowledge. However, in general we will be uncertain of at least
some of those parameters and functions. We can set those parameters and functions randomly.
Or alternatively, as described below, we can set them adversarially. It all depends on what precise
questions concerning the co-evolution of the Society and Environment agents were interested in.

\item However, the situation with the dynamics functions and allocation distribution of the Society agent  is more
complicated. Indeed, simply by allowing the values of the computation parameters
of the Society agent to change from one iteration to the next means that the \textit{forms} of
the dynamics functions of that agent will be different in different iterations. In particular,
changing the computation parameters in general changes the very spaces
that the dynamics functions of the Society agent are defined over.
(For example, if the number of machines grows, both the state spaces of the new machines and the
message graph connecting them needs to change.)

\item Like many of the other details of our model, this is a perfect place to introduce domain knowledge. 
In particular, see the discussion below of Guttman scaling for how domain knowledge might lead us to restrict
the allowed sequences of what kinds of new technology machines can exist in the society 
as we go from one iteration to the next. But we could also specify some or all of these details based on purely mathematical approaches.

\item One such purely mathematical approach is to simply generate the dynamics functions and allocation distribution of the Society 
agent in an iteration randomly. We could then
investigate how the co-evolution of the Society and Environment agents depends on initial values of the computation parameters
of the two agents, at the beginning of the first iteration, or how it depends on the 
parameters of the random distributions we are using to generate the dynamics functions. We could also investigate how the co-evolution
depends on the allocation distribution and thermodynamic cost functions of the society
agent, if we wish to consider alternative forms of those functions.

\item Another set of approaches instead involves a nested pair of loops of optimizations,
with an inner loop inside an outer loop. These approaches
start by setting the dynamics functions and allocation distribution of the Society agent in the inner loop,
based on a given dynamics of the Environment agent. Specifically, in the inner loop
we (approximately) solve for the dynamics functions and allocation distribution of the society agent so that the (expected future discounted)
GFER of the Society agent is maximized, for that given dynamics of the environment agent. 

Note that this optimization in the inner loop is a constrained optimization.
Some of the constraints are the dynamics functions and allocation distribution of the Environment agent of course.
But in addition, the optimization is constrained by the given values of the computation parameters of
the two agents in the current iteration. Furthermore, as described below in Sec. 7.2.6,
we might want to impose (possibly severe) constraints on the possible update functions of certain ``resource store'' machines, 
and on the update functions of any other machines that send messages to (or receive messages from) such
resource store machines.

Also, we need to be careful in this optimization
to respect the precise spaces that we're optimizing over.
In particular, the update functions will be functions from a finite set to itself, so the optimization
of those functions must be over some parametric form of functions from a finite set to itself.
Domain knowledge which leads us to specify the physical quantity a given machine is supposed to represent
could help here. For example, suppose we know that the elements of $X_v$ for some machine $v$
have numeric meaning, allowing us to identify them with the smallest counting numbers, $\{0, 1, 2, \ldots, |X_v| - 1 \}$.
This set is just a ring, if we take addition and multiplication to be modulo $|X_v| - 1$ (in the formal, algebraic sense
of the word). Accordingly we might want to restrict the optimization
of $f_v$ to considering only the set of linear functions over that ring. As another possibility, if $X_v$ is purely
categorical, then we may need to parameterize $f_v$ by the set of all functions from $X_v$ into itself.

A variant of this inner loop optimization is to have some of the Society agent's dynamics functions and allocation distribution 
chosen randomly, while the 
others are optimized. For example, we could have the
message graph chosen randomly, e.g.,, by sampling a Renyi process, based on the optimized value of the computation parameter
$D(G^{S})$.

In the outer loop of this nested pair of loops we need to specify the dynamics functions and allocation distribution of the Environment agent.
Indeed, until we specify how this outer loop is performed, we have not fully specified how to do the inner
loop optimization, since that inner loops needs to be given the dynamics functions and allocation distribution of the environment agent. 

There are also several ways we could do this outer loop. The first is to fix the update functions and message graph
of the environment by randomly sampling over the set of possible functions $\{f^{E}_v\}$ and graphs $G^{E}$
of the machines in the environment. (Presumably this would require us to specify a set of possible forms
of those functions and that graph, e.g., requiring the first to be linear if the states of the machines are all rings
requiring the second to be a Renyi graph for some appropriate edge probability parameter.)

A second approach is to do a full minimax calculation. In this approach, the update functions,
allocation distribution and message graph of the Society agent
are optimized in an inner loop based on a given Environment agent,
as described above --- and then the update functions and message graph of the environment
agent is adversarially set, to make the resultant (expected future discounted) GFER of the Society agent be as small
as possible.

Note though that if we adopt this second approach of analyzing the worst-case behavior of the environment in
the outer loop, then at least part of our calculation is trivial: just choose the environment agent to ignore its input
from the society agent entirely. So we have to quantify a minimal value for that dependence somehow. E.g., 
as the conditional mutual information between the environment's output at iteration $t+1$ and its input
from the society agent in that iteration (i.e., the output of the society agent at the end of iteration $t$),
conditioned on the earlier state of the environment's output, at the end of iteration $t$. (Or perhaps
in terms of conditional entropy rather than conditional mutual information.) 

\item It is important to re-emphasize that there is no right or wrong choice among the various approaches
described above for how to evolve the computation parameters and dynamics functions and allocation distribution from one iteration
to the next. It all depends on what precise question concerning the co-evolution of the Society and Environment
agents we are interested in.

\end{enumerate}

\section{Appendix B: Mathematical features of the
evolution of the Society agent across iterations}
\label{app:A}

In this appendix we briefly some of the challenges and strengths we foresee in trying
to pursue the calculations described in~\cref{sec:closing_loop} .

\begin{enumerate}

\item Note that in standard ``approximation complexity theory'' studied in CS, we are provided a desired
family of functions $\{g_i : \Sigma^*
\rightarrow \Sigma^*\}$ indexed by the length $i$ of an ``input''. (As usual, $\Sigma$ is a fixed alphabet being 
used to encode both the input and output of $g$ as strings.) Typically the associated output is viewed as encoding a scalar number.
We are concerned with the number of timesteps it
takes a computer (typically a TM) to produce an output that lies within some ``approximation ratio'' $\alpha$ of $g(x)$ for an input $x$ (usually
in the worst-case sense). The central concern, as in (almost) all of computational complexity theory, is how that number
of required timesteps scales with the size of the input $x$. (See also previous literature on ``the hardness of
approximation'' --- see~\cite{arora2009computational}, and also the work on time-bounded Kolmogorov complexity involving approximate
inference discussed in~\cite{trakhtenbrot1984survey}.) 

In contrast, we will be concerned 
with a different scaling behavior from that considered in conventional approximation complexity theory: 
For a fixed maximal number of timesteps, $\tau$, 
how does $\alpha$ scale with the size of the input? 

\item Since the Society agent only ever has limited information concerning the state of the variables in the Environment agent,
then \textit{as far
as the Society agent is concerned}, the Environment agent is a hidden Markov model (HMM). 

If we restrict the conditional distribution  $P(e^B_v(t) \,|\, x^A(t-1))$ appropriately, then even if each
variable $v \in V^B$ in the Environment receives messages directly from the same 
set of machines in the Society, there are
some such machines (occupation specifically) in the Society whose choices have no \textit{direct} impact on 
the Environment. (For example, the choices of a king have no direct impact; their only
effect on the Environment is indirect, through the actions of the \textit{other} occupations that the
king commands.) In this case, there is the same hidden nature  
in how the Society agent appears to the  Environment agent as vice-versa. 
So in this case, we actually have two interacting HMMs.

\item If in fact our two agents are two interacting HMMs, it is non-trivial (formally impossible in fact?) for the Society
agent to exactly emulate the Environment agent, even if it had access to the first input to the Environment agent, at the beginning
of the very first iteration. Intuitively,
since the Society agent is concerned with expected discounted future free energy gain (or some such), this means that the Society 
agent is making stochastic predictions for the future states of the Environment agent, and in particular of how
the future outputs of that Environment agent would differ depending on the current outputs of the society agent.


\item Now it's common to consider MDPs (Markov decision processes) where an agent is continually interacting
with an environment that is evolving according to a (1st order) Markov chain. 
This has had some limited success. But of course that success depends on the assumption
that there are no hidden variables in the environment, i.e., that $\Pi^{AB} = V^B$.

Regardless, people often sort-of motivate
the use of predictive coding in the context of MDPs, e.g., in theoretical neuroscience, in information-theoretic
terms, as a way to minimize expected codeword length. In the more abstract
scenarios this leads people to consider the use of things like (the sizes of) jpeg compressions or Lempel Ziv to quantify the 
required complexity of the predictive coding. 
%
Assuming 
that in our model the society MCM would in fact need to predict the dependence of the future
outputs of the environment MCM on the outputs of that society, the need for predictive coding 
would also arise, endogenously, in our model.

\item On the other hand, suppose that we assign
an optimization function to the dynamics of the environment (e.g., if follows a physics-like minimum
action principle). Also assume the joint behavior
of the two agents is a Nash equilibrium of their respective optimization functions.
Then so long as the set of iterations is infinite, there are scenarios
where \textit{even if there are no hidden variables}, the joint behavior does not just
have high complexity --- it is uncomputable in fact.
(See ~\cite{dargaj2020complete,dargaj2022discounted} and references therein.)
TBD if there would be similar phenomena in our scenario.
\end{enumerate}

\section{Appendix C: A possible major issue left out of our analysis}
\label{sec:kelly_and_beyond}

Since we are considering a sequence of (very) many iterations, and since a society will die out if its population
gets too small, we have an infinite ``cost of ruin'' possibility we need to worry about. In addition, the
harvest function might not only depend additively on the joint state of the variables in the two agents.
Instead, the harvest function might depend \textit{multiplicatively} on some of the relationship between the
two sets of variables.
For example, it might be that the relationship between a variable in the society agent and an associated
variable in the environment agent is {multiplied} 
by some function depending on the joint state of the other variables in the MCM's in order to determine the GFER of an iteration.

So it might make sense to try to incorporate the framework of Kelly gambling~\cite{coth91}, for example
into the harvest function. Indeed, the Kelly gambling  (``bet-hedging'')  framework
has been found to be quite important in understanding biological evolution~\cite{donaldson2010fitness}. More precisely, some
have argued that that framework should be extended to account for important
features of real biology~\cite{rivoire2011value,xue2017bet}. The same might also be
true if we want to incorporate that framework in the co-evolving MCMs framework. Indeed,
incorporating some variant of the Kelly gambling  framework might also \textit{simplify} our analysis, and perhaps tighten the
relationship between it and conventional information theory.

Recall the definition of the harvest function, which maps the
states of the environment and of the agent at the end of an iteration to the free energy harvest
by the society in that iteration. Suppose we also consider just one specific
environment behavior, in which it stochastically reacts to the combination of the action of the society and its current state (i.e., 
a fixed Bayes net for the environment, rather than a distribution over such Bayes nets, which is the case discussed above).
Suppose we similarly a fixed society behavior. 

Presumably we want to consider expected future discounted free energy harvest for such a fixed harvest
function, society behavior, and environment behavior. But in many circumstances the harvest function will
depend multiplicatively on the choice of the society, which can complicate things. An example is given by 
Kelly gambling~\cite{coth91}, i.e., the mathematics of the optimal strategy for long-term horse-betting,
or stock-betting, or genotype hedging, or any other scenario involving multiplicative payoffs to the better.

Concretely, suppose that there is an IID real-valued \textbf{winnings} variable 
in the environment agent, which has $k$ possible states, $W$, with elements written as $w$. 
For now, make the simplifying assumption that the value of $w$ in any 
given iteration is conditionally independent of all information that the society agent
has concerning the environment at the start of that iteration. Write the associated distribution as $p(w)$. 
 
Assume as well that the society must allocate some resources over a set of $k$ states, i.e., one of its variables, $\{b_i : i \in W\}$,
is a vector over $k$ values, which for simplicity are indexed with the $k$ values of $W$. Next, suppose
that the total amount of resource that society has to allocate,
$B := \sum_i b_i$, is given by the state of the society agent at the start of the iteration. So the ``decision''
of the society agent during that iteration is to set the fraction vector, $\{\dfrac{b_i}{B}  : i \in W\}$
Suppose as well that the harvest function for a given value of $w$ and $b$ is proportional to $w b_w$. 

Then as in Kelly gambling, the behavior of the agent that is asymptotically optimal is to set $b \propto p$, independent of
the precise set $W$ of possible values of the winnings, $w$. 
(Those values of $w$ only enter in determining the ``winnings rate'' at which the expected free
energy harvest increases from one iteration to the next.)

As an example of such multiplicative payoff, we could take $b_v$ to be the number of people in occupation $v$
in the associated iteration, and have $w$ be some ``winnings'' computation parameter of the environment. The idea is that the free energy harvest 
at the end of an iteration depends multiplicatively on the number of people in the occupation corresponding to the
value of that winnings computation parameter of the environment had in that iteration.  
See~\cite{donaldson2010fitness} and other papers in the population biology
literature on the ``fitness value of information''.

N.b., if the value of $w_i$ is \textit{not}  conditionally independent of all information that the society agent
has concerning the environment at the start of that iteration, then things are more complicated.  Let $y$ be the information the society agent has that is relevant to the value of $w_i$  at the beginning of
an iteration. For simplicity, suppose $y$ is a noisy observation of $w$
In this case, the optimal resource allocation changes to $b_w \propto P(w \,|\, y)$, and the associated
winnings rate increases by the mutual information between $w$ and $y$.

Of course, the harvest functions we are going to be interested in will not have \textit{exactly} multiplicative
payoffs. Nor are we interested in asymptotically optimal behavior; the number of iterations into the
future we are interested in will presumably be some finite number, and so if it's nonetheless large, we must use
deviation theory~\cite{touchette2009large} to analyze optimal behavior. These kinds of complications mean we may need to 
extend Kelly gambling somehow.

\bibliographystyle{amsplain}
\bibliography{/Users/davidwolpert/Dropbox/BIB/refs.main.1.BIB.DIR}

\end{document}